\documentclass[12pt]{article}
\usepackage{epsfig}
\usepackage{psfrag}
\usepackage{amsmath}
\usepackage{hhline}
\usepackage{amssymb}
\usepackage{times}
\usepackage{cite}
\usepackage{setspace}

\newlength{\dinwidth}
\newlength{\dinmargin}
\newcommand{\Rp}{\mbox{$\not \hspace{-0.15cm} R_p$}}
\newcommand{\lsim}{\raisebox{-1.5mm}{$\:\stackrel{\textstyle{<}}{\textstyle{\sim}}\:$}}
\newcommand{\gsim}{\raisebox{-0.5mm}{$\stackrel{>}{\scriptstyle{\sim}}$}}
\newcommand{\GeV}{\rm GeV}

\begin{document}  
\newcommand{\pom}{{I\!\!P}}
\newcommand{\reg}{{I\!\!R}}
\newcommand{\slowpi}{\pi_{\mathit{slow}}}
\newcommand{\fiidiii}{F_2^{D(3)}}
\newcommand{\fiidiiiarg}{\fiidiii\,(\beta,\,Q^2,\,x)}
\newcommand{\n}{1.19\pm 0.06 (stat.) \pm0.07 (syst.)}
\newcommand{\nz}{1.30\pm 0.08 (stat.)^{+0.08}_{-0.14} (syst.)}
\newcommand{\fiidiiiful}{F_2^{D(4)}\,(\beta,\,Q^2,\,x,\,t)}
\newcommand{\fiipom}{\tilde F_2^D}
\newcommand{\ALPHA}{1.10\pm0.03 (stat.) \pm0.04 (syst.)}
\newcommand{\ALPHAZ}{1.15\pm0.04 (stat.)^{+0.04}_{-0.07} (syst.)}
\newcommand{\fiipomarg}{\fiipom\,(\beta,\,Q^2)}
\newcommand{\pomflux}{f_{\pom / p}}
\newcommand{\nxpom}{1.19\pm 0.06 (stat.) \pm0.07 (syst.)}
\newcommand {\gapprox}
   {\raisebox{-0.7ex}{$\stackrel {\textstyle>}{\sim}$}}
\newcommand {\lapprox}
   {\raisebox{-0.7ex}{$\stackrel {\textstyle<}{\sim}$}}
\def\gsim{\,\lower.25ex\hbox{$\scriptstyle\sim$}\kern-1.30ex%
\raise 0.55ex\hbox{$\scriptstyle >$}\,}
\def\lsim{\,\lower.25ex\hbox{$\scriptstyle\sim$}\kern-1.30ex%
\raise 0.55ex\hbox{$\scriptstyle <$}\,}
\newcommand{\pomfluxarg}{f_{\pom / p}\,(x_\pom)}
\newcommand{\dsf}{\mbox{$F_2^{D(3)}$}}
\newcommand{\dsfva}{\mbox{$F_2^{D(3)}(\beta,Q^2,x_{I\!\!P})$}}
\newcommand{\dsfvb}{\mbox{$F_2^{D(3)}(\beta,Q^2,x)$}}
\newcommand{\dsfpom}{$F_2^{I\!\!P}$}
\newcommand{\gap}{\stackrel{>}{\sim}}
\newcommand{\lap}{\stackrel{<}{\sim}}
\newcommand{\fem}{$F_2^{em}$}
\newcommand{\tsnmp}{$\tilde{\sigma}_{NC}(e^{\mp})$}
\newcommand{\tsnm}{$\tilde{\sigma}_{NC}(e^-)$}
\newcommand{\tsnp}{$\tilde{\sigma}_{NC}(e^+)$}
\newcommand{\st}{$\star$}
\newcommand{\sst}{$\star \star$}
\newcommand{\ssst}{$\star \star \star$}
\newcommand{\sssst}{$\star \star \star \star$}
\newcommand{\tw}{\theta_W}
\newcommand{\sw}{\sin{\theta_W}}
\newcommand{\cw}{\cos{\theta_W}}
\newcommand{\sww}{\sin^2{\theta_W}}
\newcommand{\cww}{\cos^2{\theta_W}}
\newcommand{\trm}{m_{\perp}}
\newcommand{\trp}{p_{\perp}}
\newcommand{\trmm}{m_{\perp}^2}
\newcommand{\trpp}{p_{\perp}^2}
\newcommand{\alp}{\alpha_s}

\newcommand{\alps}{\alpha_s}
\newcommand{\sqrts}{$\sqrt{s}$}
\newcommand{\LO}{$O(\alpha_s^0)$}
\newcommand{\Oa}{$O(\alpha_s)$}
\newcommand{\Oaa}{$O(\alpha_s^2)$}
\newcommand{\PT}{p_{\perp}}
\newcommand{\JPSI}{J/\psi}
\newcommand{\sh}{\hat{s}}
\newcommand{\uh}{\hat{u}}
\newcommand{\MP}{m_{J/\psi}}
\newcommand{\PO}{I\!\!P}
\newcommand{\xbj}{x}
\newcommand{\xpom}{x_{\PO}}
\newcommand{\ttbs}{\char'134}
\newcommand{\xpomlo}{3\times10^{-4}}  
\newcommand{\xpomup}{0.05}  
\newcommand{\dgr}{^\circ}
\newcommand{\pbarnt}{\,\mbox{{\rm pb$^{-1}$}}}
\newcommand{\gev}{\,\mbox{GeV}}
\newcommand{\WBoson}{\mbox{$W$}}
\newcommand{\fbarn}{\,\mbox{{\rm fb}}}
\newcommand{\fbarnt}{\,\mbox{{\rm fb$^{-1}$}}}
%
%
\newcommand{\qsq}{\ensuremath{Q^2} }
\newcommand{\gevsq}{\ensuremath{\mathrm{GeV}^2} }
\newcommand{\et}{\ensuremath{E_t^*} }
\newcommand{\rap}{\ensuremath{\eta^*} }
\newcommand{\gp}{\ensuremath{\gamma^*}p }
\newcommand{\dsiget}{\ensuremath{{\rm d}\sigma_{ep}/{\rm d}E_t^*} }
\newcommand{\dsigrap}{\ensuremath{{\rm d}\sigma_{ep}/{\rm d}\eta^*} }
\def\Journal#1#2#3#4{{#1} {\bf #2} (#3) #4}
\def\NCA{\em Nuovo Cimento}
\def\NIM{\em Nucl. Instrum. Methods}
\def\NIMA{{\em Nucl. Instrum. Methods} {\bf A}}
\def\NPB{{\em Nucl. Phys.}   {\bf B}}
\def\PLB{{\em Phys. Lett.}   {\bf B}}
\def\PRL{\em Phys. Rev. Lett.}
\def\PRD{{\em Phys. Rev.}    {\bf D}}
\def\ZPC{{\em Z. Phys.}      {\bf C}}
\def\EJC{{\em Eur. Phys. J.} {\bf C}}
\def\CPC{\em Comp. Phys. Commun.}

\begin{titlepage}

\begin{figure}[!t]
DESY 04--025 \hfill ISSN 0418--9833\\ 
March 2004 
\end{figure}
\bigskip

\vspace*{2cm}

\begin{center}
\begin{Large}

{\bf Search for Squark Production\\ in R-Parity Violating Supersymmetry at HERA}

\vspace{2cm}

H1 Collaboration

\end{Large}
\end{center}

\vspace{2cm}

\begin{abstract}
A search for squarks in $R$-parity violating supersymmetry is performed 
in $e^{\pm} p$ collisions at HERA using the H1 detector.
The data were taken at a centre-of-mass energy of $319\,\GeV$ and correspond 
to an integrated luminosity of $64.3\,\mathrm{pb^{-1}}$ for $e^+p$ collisions 
and $13.5\,\mathrm{pb^{-1}}$ for $e^-p$ collisions.
The resonant production of squarks via a Yukawa coupling $\lambda'$ is considered, 
taking into account direct and indirect $R$-parity violating decay modes.
No evidence for squark production is found in the multi-lepton and multi-jet 
final state topologies investigated.
Mass dependent limits on $\lambda'$ are obtained in the framework of the Minimal 
Supersymmetric Standard Model.
In addition, the results are interpreted in terms of constraints on the parameters 
of the minimal Supergravity model. 
At the $95\,\%$ confidence level squarks of all flavours with masses up to 
$275\,\GeV$ are excluded in a large part of the parameter space for a Yukawa 
coupling of electromagnetic strength.
For a coupling strength 100 times smaller, masses up to $220\,\GeV$ can be 
ruled out.
\end{abstract}

\vspace{1.5cm}

\begin{center}
To be submitted to {\it Eur.\ Phys.\ J.\ C.} 
\end{center}

\end{titlepage}

\begin{flushleft}

A.~Aktas$^{10}$,               
V.~Andreev$^{26}$,             
T.~Anthonis$^{4}$,             
A.~Asmone$^{33}$,              
A.~Babaev$^{25}$,              
S.~Backovic$^{37}$,            
J.~B\"ahr$^{37}$,              
P.~Baranov$^{26}$,             
E.~Barrelet$^{30}$,            
W.~Bartel$^{10}$
% ...

\bigskip{\it
% ...
$ ^{4}$ Inter-University Institute for High Energies ULB-VUB, Brussels;
          Universiteit Antwerpen (UIA), Antwerpen; Belgium$^{ c}$ \\
% ...
$ ^{10}$ DESY, Hamburg, Germany \\
% ...
$ ^{25}$ Institute for Theoretical and Experimental Physics,
          Moscow, Russia$^{ l}$ \\
 $ ^{26}$ Lebedev Physical Institute, Moscow, Russia$^{ e}$ \\
% ...
$ ^{30}$ LPNHE, Universit\'{e}s Paris VI and VII, IN2P3-CNRS,
          Paris, France \\
% ...
$ ^{33}$ Dipartimento di Fisica Universit\`a di Roma Tre
          and INFN Roma~3, Roma, Italy \\
% ...
$ ^{37}$ DESY, Zeuthen, Germany \\
% ...
$ ^c$ Supported by FNRS-FWO-Vlaanderen, IISN-IIKW and IWT \\
% ...
$ ^e$ Supported by the Deutsche Forschungsgemeinschaft \\
% ...
$ ^l$ Partially Supported by Russian Foundation
      for Basic Research, grant    no. 00-15-96584 \\
}

\end{flushleft}

\newpage
\section{Introduction}
The $ep$ collider HERA is ideally suited to search for new particles coupling to 
electron\footnote{In the following the term {\it electron} will be used to 
refer to both electron and positron unless explicitly stated otherwise.}--quark 
pairs.
In supersymmetric models (SUSY) with $R$-parity violation (\Rp ), squarks 
can couple to electrons and quarks via Yukawa couplings $\lambda'$.
At HERA, squarks could be produced resonantly via the fusion of the incoming 
$27.6\,\GeV$ electron and a quark from the incoming $920\,\GeV$ proton.
Squark masses up to the electron-proton centre-of-mass energy, $\sqrt{s}=319\,\GeV$, 
are kinematically accessible.

This paper describes a search for squarks of all flavours using 
H1 data corresponding 
to an integrated luminosity of $64.3\,\mathrm{pb^{-1}}$ for $e^+p$ collisions and 
$13.5\,\mathrm{pb^{-1}}$ for $e^-p$ collisions.
The search is carried out in the framework of the Minimal Supersymmetric Standard 
Model (MSSM) in the presence of a non-vanishing $\lambda'$.
The analysis covers the major event topologies from direct and indirect \Rp\ squark 
decays such that the results can be interpreted in 
terms of a wide range of SUSY parameters.
The search presented here supersedes the results previously obtained by 
H1~\cite{PEREZ,OLDLIMIT} at a lower centre-of-mass energy ($\sqrt{s}\approx 300\,\GeV$) 
and with fewer data.
Complementary searches for \Rp\  SUSY have been carried out at the LEP $e^+e^-$ 
collider~\cite{LEP,L3RPV} and at the TeVatron $p\bar{p}$ collider~\cite{TEVATRON,D0RES}. 

\section{Phenomenology}
In the most general supersymmetric theory that is renormalisable and gauge invariant 
with respect to the Standard Model (SM) gauge group, the $R$-parity 
$R_p=(-1)^{3B+L+2S}$, where $B$ denotes the baryon number, $L$ the lepton number 
and $S$ the spin of a particle, is not conserved.
Couplings between two ordinary fermions and a squark ($\tilde{q}$) or a 
slepton ($\tilde{l}$) are then allowed.
The \Rp\ Yukawa couplings responsible for squark production at 
HERA are described in the superpotential by the terms 
$\lambda'_{ijk} L_{i}Q_{j}\overline{D}_k$, where $i,j$ and $k$ are family indices.
$L_i$, $Q_j$ and $D_k$ are superfields, which contain the left-handed leptons, 
the left-handed quarks and the right-handed down quark, respectively, together 
with their SUSY partners $\tilde{l}_L^i$, $\tilde{q}_L^j$ and $\tilde{d}_R^k$.
The corresponding part of the Lagrangian expanded in fields is given by
\begin{eqnarray}
{\cal{L}}_{L_{i}Q_{j}\overline{D}_{k}} &=
   & \lambda^{\prime}_{ijk}
              \left[ -\tilde{e}_{L}^{i} u^j_L \bar{d}_R^k
              - e^i_L \tilde{u}^j_L \bar{d}^k_R - (\bar{e}_L^i)^c u^j_L
     \tilde{d}^{k*}_R \right.           \nonumber \\
 \mbox{} &\mbox{}
 & \left. + \tilde{\nu}^i_L d^j_L \bar{d}^k_R + \nu_L \tilde{d}^j_L
    \bar{d}^k_R + (\bar{\nu}^i_L)^c d^j_L \tilde{d}^{k*}_R \right]
   +\mbox{c.c.},             
 \label{eq:lagrangian}
\end{eqnarray}
where the superscript $c$ denotes the charge conjugate of a spinor
and $*$ the complex conjugate of a scalar field.
Non-vanishing couplings $\lambda'_{1jk}$ allow the resonant production of 
squarks at HERA through $eq$ fusion~\cite{RPVIOLATION}.
The values of the couplings are not fixed by the theory.
For simplicity, it is assumed here that one of the $\lambda'_{1jk}$ 
dominates over all other possible trilinear couplings.

For the nine possible couplings $\lambda'_{1jk}$, the corresponding squark 
production processes in $e^{\pm}p$ reactions are listed in table~\ref{tab:sqprod}.
At high Bjorken-$x$ the density of antiquarks in the proton is smaller than that of quarks.
Thus $e^-p$ scattering gives the best sensitivity to the couplings $\lambda'_{11k}$ 
($k=1,2,3$), where mainly $\tilde{d}_R$-type squarks ({\it i.e.} the superpartners 
$\tilde{d}_R,\tilde{s}_R$ and $\tilde{b}_R$ of right-handed quarks) can be produced.
The dominant squark production cross section in $e^-p$ collisions is approximately 
proportional to {\mbox{$\lambda^{'2}_{11k} \cdot u(x)$}} where $u(x)$ is the 
probability of finding a $u$ quark in the proton with a momentum fraction 
$x=M^2_{\tilde{q}}/s$, $M_{\tilde{q}}$ being the squark mass.
In contrast, $e^+p$ scattering gives the best sensitivity to the couplings 
$\lambda'_{1j1}$ ($j=1,2,3$), where mainly $\tilde{u}_L$-type squarks ({\it i.e.} 
the superpartners $\tilde{u}_L,\tilde{c}_L$ and $\tilde{t}_L$ of right-handed 
quarks) can be produced.
Here the dominant squark production cross section is approximately proportional to 
{\mbox{$\lambda^{'2}_{1j1} \cdot d(x)$}}.
Since the $u$ quark density in the proton is larger than the $d$ quark density
at large $x$,
the squark production cross section in $e^-p$ 
interactions is larger than that in the $e^+p$ case
for comparable couplings .

In this work the signal cross section is obtained in the narrow width approximation 
from the leading order (LO) amplitudes given in \cite{BUCHMULL}, corrected by 
multiplicative factors~\cite{SPIRANLO} to account for next-to-leading order QCD corrections. 
The parton densities are evaluated at the hard scale $M_{\tilde{q}}^2$.
For cases, where the squark width is not negligible, the approach 
given in\cite{MYTHESIS} is followed.

%
%
\begin{table*}[t]
 \begin{center}
   \begin{tabular}{|c||c|c||c|c|}
   \hline
    $\lambda'_{1jk}$ & \multicolumn{2}{c||}{\boldmath $e^-p$} & \multicolumn{2}{c|}{\boldmath $e^+p$} \\ \hline
\hline \rule[0mm]{0mm}{5mm}
   111 & 
$e^- +u \rightarrow \tilde{d}_R$ & $e^- +\overline{d} \rightarrow \overline{\tilde{u}}_L$ &
$e^+ +d \rightarrow \tilde{u}_L $& $e^+ +\overline{u} \rightarrow \overline{\tilde{d}}_R$ \\

   112 & 
$e^- +u \rightarrow \tilde{s}_R$ & $e^- +\overline{s} \rightarrow \overline{\tilde{u}}_L$ &
$e^+ +s \rightarrow \tilde{u}_L $ & $e^+ +\overline{u} \rightarrow \overline{\tilde{s}}_R$ \\

   113 & 
$e^- +u \rightarrow \tilde{b}_R$ & $e^- +\overline{b} \rightarrow \overline{\tilde{u}}_L$ &
$e^+ +b \rightarrow \tilde{u}_L $ & $e^+ +\overline{u} \rightarrow \overline{\tilde{b}}_R$ \\

   121 & 
$e^- +c \rightarrow \tilde{d}_R$ & $e^- +\overline{d} \rightarrow \overline{\tilde{c}}_L$ &
$e^+ +d \rightarrow \tilde{c}_L $ & $e^+ +\overline{c} \rightarrow \overline{\tilde{d}}_R$ \\

   122 & 
$e^- +c \rightarrow \tilde{s}_R$ & $e^- +\overline{s} \rightarrow \overline{\tilde{c}}_L$ &
$e^+ +s \rightarrow \tilde{c}_L $ & $e^+ +\overline{c} \rightarrow \overline{\tilde{s}}_R$ \\

   123 & 
$e^- +c \rightarrow \tilde{b}_R$ & $e^- +\overline{b} \rightarrow \overline{\tilde{c}}_L$ &
$e^+ +b \rightarrow \tilde{c}_L $ & $e^+ +\overline{c} \rightarrow \overline{\tilde{b}}_R$ \\

   131 & 
$e^- +t \rightarrow \tilde{d}_R$ & $e^- +\overline{d} \rightarrow \overline{\tilde{t}}_L$ &
$e^+ +d \rightarrow \tilde{t}_L $ & $e^+ +\overline{t} \rightarrow \overline{\tilde{d}}_R$ \\

   132 & 
$e^- +t \rightarrow \tilde{s}_R$ & $e^- +\overline{s} \rightarrow \overline{\tilde{t}}_L$ &
$e^+ +s \rightarrow \tilde{t}_L $ & $e^+ +\overline{t} \rightarrow \overline{\tilde{s}}_R$ \\

   133 & 
$e^- +t \rightarrow \tilde{b}_R$ & $e^- +\overline{b} \rightarrow \overline{\tilde{t}}_L$ & 
$e^+ +b \rightarrow \tilde{t}_L $ & $e^+ +\overline{t} \rightarrow \overline{\tilde{b}}_R$ \\
   \hline
  \end{tabular}

\end{center}
  \caption{Squark production processes in $e^{\pm}p$ collisions for 
different Yukawa couplings $\lambda'_{1jk}$. The $\tilde{q}_R$ and $\tilde{q}_L$ 
symbols denote the squarks which are superpartners of the right- and left-handed 
quarks, respectively. Their antimatter counterparts are denoted by 
$\overline{\tilde{q}}_R$ and $\overline{\tilde{q}}_L$.}
\label{tab:sqprod}
\end{table*}
%
\begin{figure}[b]
 \begin{center}
\psfrag{a}[][][1.0][0]{$e^-$}
\psfrag{b}[][][1.0][0]{$u$}
\psfrag{c}[][][1.0][0]{$\tilde{d}_R^k$}
\psfrag{d}[][][1.0][0]{\quad $e^-,\nu_e$}
\psfrag{e}[][][1.0][0]{\quad $u,d$}
\psfrag{l1}[][][1.0][0]{$\lambda'_{11k}$}
\psfrag{l2}[][][1.0][0]{$\lambda'_{11k}$}
\psfrag{N}[][][1.0][0]{(a)}
  \epsfig{file=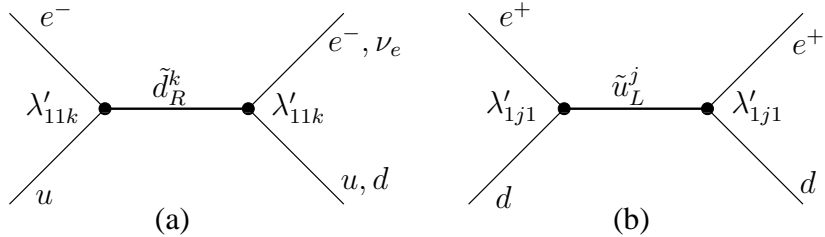, scale=0.5, angle=0.0}
\psfrag{a}[][][1.0][0]{$e^+$}
\psfrag{b}[][][1.0][0]{$d$}
\psfrag{c}[][][1.0][0]{$\tilde{u}_L^j$}
\psfrag{d}[][][1.0][0]{$e^+$}
\psfrag{e}[][][1.0][0]{$d$}
\psfrag{l1}[][][1.0][0]{$\lambda'_{1j1}$}
\psfrag{l2}[][][1.0][0]{$\lambda'_{1j1}$}
\psfrag{N}[][][1.0][0]{(b)}
\epsfig{file=paper1.eps, scale=0.5, angle=0.0}
\end{center}
   \caption{Lowest order $s$-channel diagrams for \Rp\ squark production 
via the Yukawa coupling $\lambda'$ in (a) $e^-p$ and (b) $e^+p$ interactions, 
followed by \Rp\ squark decays.}
  \label{fig:rpvdec}

\end{figure}

\begin{figure}[h]
 \begin{center}
\psfrag{a}[][][1.0][0]{$e^-$}
\psfrag{b}[][][1.0][0]{$u$}
\psfrag{c}[][][1.0][0]{$\tilde{d}_R^k$}
\psfrag{d}[][][1.0][0]{$d$}
\psfrag{e}[][][1.0][0]{$\chi_i^0,\tilde{g}$}
\psfrag{f}[][][1.0][0]{$\chi_1^0$}
\psfrag{g}[][][1.0][0]{$e^+$}
\psfrag{j}[][][1.0][0]{$\tilde{e}^-$}
\psfrag{h}[][][1.0][0]{$\bar{u}$}
\psfrag{i}[][][1.0][0]{$d^k$}
\psfrag{l1}[][][0.8][0]{$\lambda'_{11k}$}
\psfrag{l2}[][][0.8][0]{$\lambda'_{11k}$}
\psfrag{N}[][][1.0][0]{(a)}
  \epsfig{file=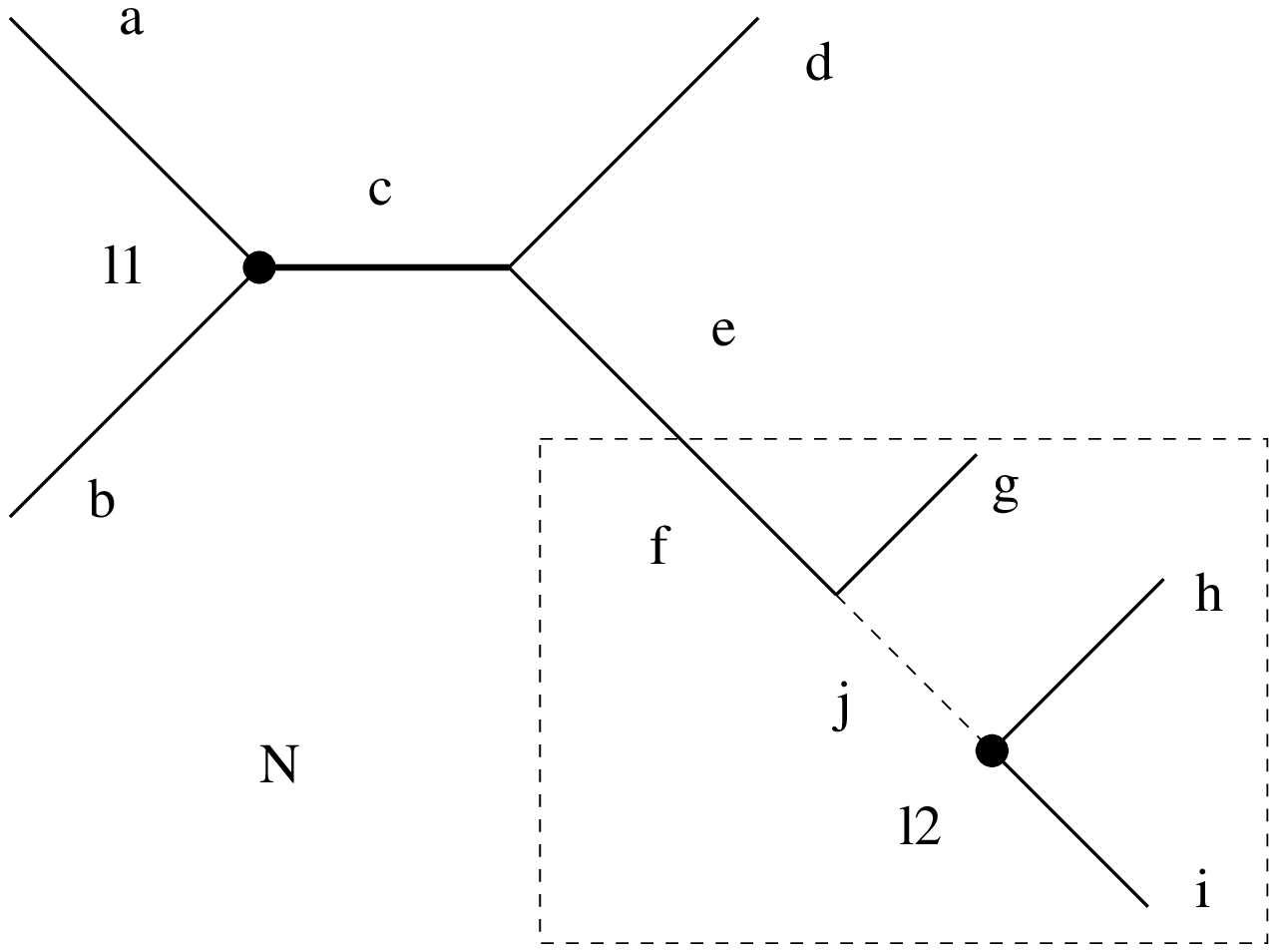, scale=0.5, angle=0.0}
\psfrag{a}[][][1.0][0]{$e^+$}
\psfrag{b}[][][1.0][0]{$d$}
\psfrag{c}[][][1.0][0]{$\tilde{u}_L^j$}
\psfrag{d}[][][1.0][0]{$u,d$}
\psfrag{e}[][][1.0][0]{$\chi_i^0,\chi^+_i,\tilde{g}$}
\psfrag{f}[][][1.0][0]{$\chi_1^+$}
\psfrag{1}[][][1.0][0]{$eq\bar{q}$}
\psfrag{2}[][][1.0][0]{$(\nu_e q \bar{q})$}
\psfrag{g}[][][1.0][0]{$\chi_1^0$}
\psfrag{j}[][][1.0][0]{$W^+$}
\psfrag{h}[][][1.0][0]{$q',l^+$}
\psfrag{i}[][][1.0][0]{$\bar{q},\nu_l$}
\psfrag{l1}[][][0.8][0]{$\lambda'_{1j1}$}
\psfrag{l2}[][][0.8][0]{$\lambda'_{1j1}$}
\psfrag{N}[][][1.0][0]{(b)}
  \epsfig{file=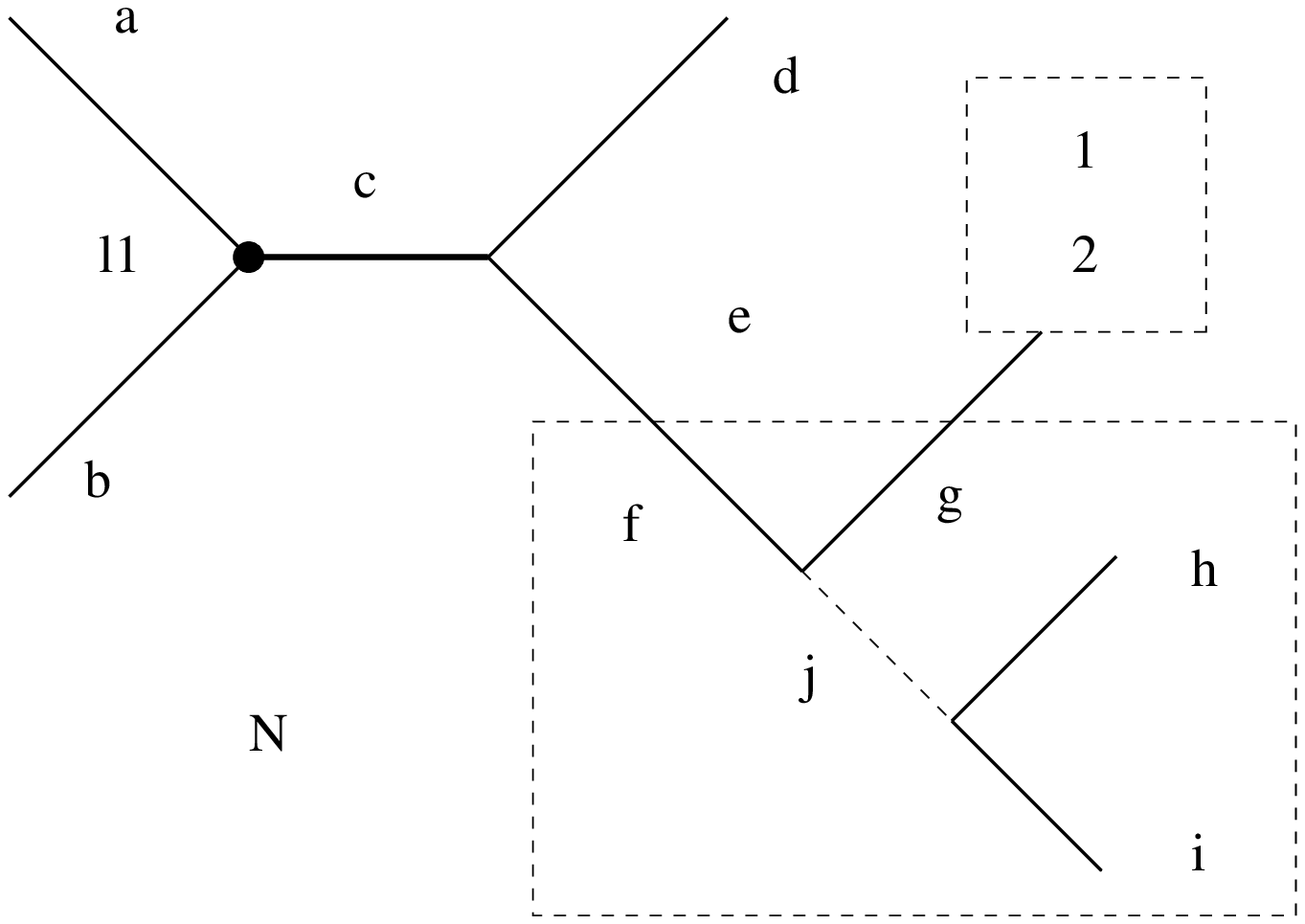, scale=0.5, angle=0.0}
 \end{center}
   \caption{Gauge decays of squarks. Example decays of the emerging neutralino, chargino or 
gluino are shown in the dashed boxes for (a) the $\chi_1^0$ and (b) the $\chi_1^+$.}
  \label{fig:gaudec}
\end{figure}

In \Rp\ SUSY all supersymmetric particles are unstable.
Squarks can decay via their Yukawa coupling $\lambda'$ into SM fermions.
According to equation~(\ref{eq:lagrangian}), the $\tilde{d}_R$-type squarks 
can decay either into $e^- + {u}^j$ or $\nu_e + {d}^j$, while the $\tilde{u}_L$-type 
squarks decay into $e^+ + d^{k}$ only.
The \Rp\ squark decays, proceeding directly via the couplings $\lambda'_{11k}$ 
and $\lambda'_{1j1}$, are illustrated in figure~\ref{fig:rpvdec}.

Squarks can also decay via their usual $R_p$ conserving gauge couplings, as 
shown in figure~\ref{fig:gaudec}. 
The $\tilde{u}_L$-type squarks can undergo a gauge decay into states\footnote{The 
mass eigenstates $\chi^0_i$ ($\chi^{\pm}_i$) are mixed states of the photino, the 
zino and the neutral higgsinos (the winos and the charged higgsinos). The $\tilde{g}$ 
is the SUSY partner of the gluon.} involving a neutralino $\chi^0_i$ 
($i=1\dots 4$), a chargino $\chi^{+}_i$ ($i=1,2$) or a gluino $\tilde{g}$.
In contrast, $\tilde{d}_R$-type squarks decay to $\chi^0_i$ or $\tilde{g}$ only 
and decays into charginos are suppressed, since the supersymmetric partners of 
right-handed quarks do not couple to winos.

The final state of these gauge decays depends on the subsequent gaugino decay, 
of which examples are shown in figure~\ref{fig:gaudec}.
Neutralinos $\chi_i^0$ with $i>1$ as well as charginos (gluinos) are expected 
to undergo gauge decays into a lighter $\chi$ and two SM fermions (two quarks),
through a real or virtual gauge boson or sfermion (squark).
The decay chain ends with the \Rp\ decay of one sparticle, usually the lightest 
supersymmetric particle (LSP), assumed here to be a $\chi^0$, $\chi^{\pm}$ or 
$\tilde{g}$.
\Rp\ decays of gauginos are mainly relevant for the lightest states.
Neutralinos may undergo the \Rp\ decays $\chi^0 \rightarrow e^{\pm} q \bar{q}'$ or
$\chi^0 \rightarrow \nu q \bar{q}$, the former (latter) being more frequent if 
the $\chi^0$ is dominated by its photino (zino) component.
Gluinos can undergo the same \Rp\ decays.
When a $\chi^0$ or a $\tilde{g}$ decays via an \Rp\ coupling into a charged 
lepton, both the ``right'' and the ``wrong'' charge lepton (with respect to 
the incident beam) are equally probable, the latter case leading to striking, 
largely background-free signatures for lepton number violation.
In contrast, the only possible \Rp\ decays for charginos are 
$\chi^+ \rightarrow \bar{\nu} u^k \bar{d}^j$ and $\chi^+ \rightarrow e^+ d^k \bar{d}^j$.

The $\tilde{u}^j_L$ ($\tilde{d}^k_R$) decay chains analysed in this paper
are classified by event topology, as described in table~\ref{tab:sqtopo1}.
This classification relies on the number of charged leptons and/or 
hadronic jets
in the final state, and on the presence of missing momentum.
The channels labelled $eq$ and $\nu q$ are the squark decay modes 
which proceed 
directly via \Rp\ couplings, while the remaining channels result from the gauge 
decays of the squark and are characterised by multijet (MJ) final states.
The channels 
labelled $e^+M\!J$, $e^- M\!J$ and $\nu M\!J$ involve one or two SUSY 
fermions ($\chi$ or $\tilde{g}$) denoted by $X$ and $Y$ in table~\ref{tab:sqtopo1}.
The channels $e \ell M\!J$ and $\nu \ell M\!J$ necessarily involve two SUSY fermions.

Decay patterns involving more than two gauginos are kinematically suppressed and are 
therefore not explicitly studied here.
Processes leading to final states with tau leptons are also not explicitly investigated.
Cases where the $\chi^0_1$ has such a long lifetime that large displaced vertices are 
expected are not considered, since the region of parameter space that allows a 
$\chi^0_1$ to escape detection for a finite value of the \Rp\ coupling
is very strongly constrained by the searches for gauginos carried out at LEP~\cite{L3RPV}.
Decays of a $\chi$ into states involving a Higgs boson are taken into account when the 
Higgs decays into hadrons.
The contribution of these decays is, however, very small.

\begin{table}[p]
 \renewcommand{\doublerulesep}{0.4pt}
 \renewcommand{\arraystretch}{1.0}
 \begin{center}
  \begin{tabular}{||c|l|c||}
  \hline \hline
  {\bf Channel}  &  \multicolumn{1}{c|}{\bf Decay process}
           & \multicolumn{1}{c||}{\bf Event topology} \\
           & \multicolumn{1}{c|}{ }
           & \multicolumn{1}{c||}{ }         \\ \hline
  {\boldmath{$eq$}} &  \begin{tabular}{cccccc}
          $\tilde{q}$ & $\stackrel{\lambda'}{\longrightarrow}$
                      & $e$   & $q$    &    &
        \end{tabular}
     &  \begin{tabular}{c}
        high $p_T$ $e$ + 1 jet
        \end{tabular} \\                                        \hline
  {\boldmath{$\nu q$}} &  \begin{tabular}{cccccc}

         $\tilde{d}^k_R$ & $\stackrel{\lambda'}{\longrightarrow}$
                      & $\nu_e$   & $d$     &    & 
        \end{tabular}
     &  \begin{tabular}{c}
         missing $p_T$ + 1 jet
        \end{tabular} \\                                        \hline
  {\boldmath{$e^{\pm}M\!J$}} & \begin{tabular}{ccccll}
         $\tilde{q}$ & $\longrightarrow$
                     & $q$ & $X$  &  &  \\
         &  &        & $\stackrel{\lambda'}{\hookrightarrow}$
                     & $e^{\pm} \bar{q} q$ & \\
         $\tilde{q}$ & $\longrightarrow$
                     & $q$ & $X$ &  & \\
         &  &        & $\hookrightarrow$
                     & $q \bar{q}$ & $\hspace{-0.cm}Y$ \\
         &  &  &  & 
                  & $\hspace{-0.cm}
                \stackrel{\lambda'}{\hookrightarrow}$ $e^{\pm} \bar{q} q$  \\
       \end{tabular}
     & \begin{tabular}{c}
         $e$ (both charges) \\
        + multiple jets
       \end{tabular}\\                                          \hline
  {\boldmath{$\nu M\!J$}} & \begin{tabular}{ccccll}
         $\tilde{q}$ & $\longrightarrow$ & $q$ & $X$ &  & \\
         &  &        & $\stackrel{\lambda'}{\hookrightarrow}$
         & $\nu  \bar{q} q$\hspace{0.22cm} &  \\
         $\tilde{q}$ & $\longrightarrow$ & $q$ & $X$ &  & \\
         &  &        & $\hookrightarrow$
                     & $q \bar{q}$ & $\hspace{-0.cm}Y$ \\
         &  &  & &
         & $\hspace{-0.cm}
           \stackrel{\lambda'}{\hookrightarrow}$ $\nu  \bar{q} q$ \\
         $\tilde{q}$ & $\longrightarrow$ & $q$ & $X$ &  & \\
         &  &        & $\hookrightarrow$
                     & $\nu \bar{\nu}$ & $\hspace{0cm}Y$ \\
         &  &  & &
         & $\hspace{-0.cm}
           \stackrel{\lambda'}{\hookrightarrow}$ $\nu  \bar{q} q$ \\

       \end{tabular}
     & \begin{tabular}{c}
        missing $p_T$ \\
        + multiple jets
       \end{tabular}\\                                           \hline
  {\boldmath{$elM\!J$}} & \begin{tabular}{ccccll}
         $\tilde{q}$ & $\longrightarrow$ & $q$ & $X$ &  & \\
         &  &        & $\hookrightarrow$
                     & $\ell \nu_{\ell} $ & $\hspace{0cm}Y$ \\
         &  &  & &
         & $\hspace{0cm}
           \stackrel{\lambda'}{\hookrightarrow}$ $e^{\pm} \bar{q} q$ \\

         $\tilde{q}$ & $\longrightarrow$ & $q$ & $X$ &  & \\
         &  &        & $\hookrightarrow$
                     & $\ell^+ \ell^-  $ & $\hspace{0cm}Y$ \\
        &  &  & &
         & $\hspace{0cm}
           \stackrel{\lambda'}{\hookrightarrow}$ $e^{\pm} \bar{q} q$ \\

         $\tilde{q}$ & $\longrightarrow$ & $q$ & $X$ &  & \\
         &  &        & $\hookrightarrow$
                     & $e^+ e^-  $ & $\hspace{0cm}Y$ \\
         &  &  & &
         & $\hspace{0cm}
           \stackrel{\lambda'}{\hookrightarrow}$ $\nu \bar{q} q$ 
       \end{tabular}
     & \begin{tabular}{c}
         $e$ \\
        +  $\ell$ ($e$ or $\mu$) \\
        + multiple jets
       \end{tabular}\\                                          \hline
  {\boldmath{$\nu lM\!J$}} & \begin{tabular}{ccccll}
         $\tilde{q}$ & $\longrightarrow$ & $q$ & $X$ &  & \\
         &  &        & $\hookrightarrow$
                     & $\ell \nu_{\ell} $ & $\hspace{0cm}Y$ \\
         &  &  & &
         & $\hspace{0cm}
           \stackrel{\lambda'}{\hookrightarrow}$ $\nu \bar{q} q$ \\

         $\tilde{q}$ & $\longrightarrow$ & $q$ & $X$ &  & \\
         &  &        & $\hookrightarrow$
                     & $\nu \bar{\nu} $ & $\hspace{0cm}Y$ \\
         &  &  & &
         & $\hspace{0cm}
           \stackrel{\lambda'}{\hookrightarrow}$ $e \bar{q} q$ \\
         $\tilde{q}$ & $\longrightarrow$ & $q$ & $X$ &  & \\
         &  &        & $\hookrightarrow$
                     & $\mu^+ \mu^- $ & $\hspace{0cm}Y$ \\
         &  &  & &
         & $\hspace{0cm}
           \stackrel{\lambda'}{\hookrightarrow}$ $\nu \bar{q} q$ 
       \end{tabular}
     & \begin{tabular}{c}
          $\ell$ ($e$ or $\mu$) \\
        + missing $p_T$ \\

        + multiple jets
       \end{tabular}\\                                          
   \hline \hline
  \end{tabular}
  \caption
{Squark decay channels in \Rp\ SUSY classified by event topology. 
$X$ and $Y$ denote  a neutralino, a chargino or a gluino. 
The \Rp\ process is indicated by $\lambda'$. }
  \label{tab:sqtopo1}
 \end{center}
\end{table}
\section{The H1 Detector}
A detailed description of the H1 experiment can be found in~\cite{H1DETECT}.
The main components of the tracking system are the central drift and proportional
chambers which cover the polar angle\footnote{The polar angle $\theta$ is 
measured with respect to the direction of the outgoing proton beam ($+z$).} 
range 25$^{\circ} \le \theta \le$ 155$^{\circ}$, a forward track
detector  (7$^{\circ} \le \theta \le$ 25$^{\circ}$) and a backward drift chamber.
The tracking system is surrounded by a finely segmented liquid argon (LAr) 
calorimeter~\cite{H1LARCAL} which covers the polar angle range 
4$^{\circ} \le \theta \le$ 154$^{\circ}$ and has an energy resolution of
$\sigma(E)/E \simeq$ $12\%/\sqrt{E/\GeV} \oplus1\%$ for electrons and 
$\sigma(E)/E \simeq$ $50\%/\sqrt{E/\GeV} \oplus2\%$ for hadrons, as obtained in 
test beam measurements~\cite{H1CALRES}.
The tracking chambers and the LAr are surrounded by a superconducting solenoid 
and its iron yoke instrumented with streamer tubes.
The latter are used to detect hadronic showers which extend beyond the LAr and 
to identify muons.
The luminosity is determined from the rate of Bethe-Heitler events 
($e p \rightarrow e p \gamma$) measured in a luminosity monitor. 

\section{Monte Carlo Event Generation}
\label{sec:dismc}

In order to estimate the amount of SM background in the various squark decay 
channels and to determine the signal detection efficiencies, complete 
simulations of the H1 detector response are performed for various 
Monte Carlo (MC) samples.

For each possible SM background source, a sample of MC events is used, corresponding 
to a luminosity of more than ten times that of the data.  
The determination of the contribution of neutral current (NC) deep inelastic 
scattering (DIS) processes is performed using two MC programs which both include 
LO QCD matrix elements but employ different models of QCD radiation.
The first is produced with the DJANGO~\cite{DJANGO} event generator, where QCD 
radiation is implemented using ARIADNE~\cite{ARIADNE}, 
based on the Colour Dipole Model (CDM)~\cite{CDM}.
This sample is chosen to estimate the NC DIS contribution in the $eq$ channel.
The second sample is generated with the program RAPGAP~\cite{RAPGAP}, where higher 
order QCD
radiation is modelled using leading-log DGLAP parton showers~\cite{DGLAP}. 
This sample is used to determine the NC DIS background in the final states with an 
electron and multiple jets, as RAPGAP gives the 
better description of this particular 
phase space domain~\cite{MYTHESIS}.
For both samples, the parton densities in the proton are taken from the CTEQ5L~\cite{CTEQ} 
parameterisation.
Hadronisation is performed in the Lund string fragmentation scheme using JETSET~\cite{JETSET}.
The modelling of the charged current (CC) DIS process is performed using the DJANGO program 
with CTEQ5L parton densities.
The direct and resolved photoproduction ($\gamma p$) of  light and heavy flavours, including 
prompt photon production, is generated using the PYTHIA~\cite{PYTHIA} program, which relies 
on first order matrix elements and uses leading-log parton showers and string fragmentation. 
The SM expectations for $ep\rightarrow eW^{\pm}X$ and $ep\rightarrow eZ^0X$ are calculated 
using EPVEC~\cite{EPVEC}. 
The LO MC simulations used to model QCD multi-jet production
give only approximate descriptions of the kinematic distributions. 
From the comparison of the distributions of multi-jet events 
between the data and the LO MC 
simulations a normalisation factor of 1.2 is derived~\cite{MYTHESIS} which is applied to the 
yield of multi-jet events predicted by RAPGAP and PYTHIA.

To allow a model independent interpretation of the results, all squark decay processes 
given in table~\ref{tab:sqtopo1} are simulated separately for a wide range of masses of 
the SUSY particles involved.
The LEGO~\cite{LEGO} event generator is used for the determination of the signal detection 
efficiencies in the $eq$ and $\nu q$ channels, whereas for the gauge decays of squarks 
the SUSYGEN~\cite{SUSYGEN} generator is used.
The squark mass is varied from $100\,\GeV$ to $290\,\GeV$ in steps of typically $25\,\GeV$. 
For gauge decays of squarks involving a gaugino which decays directly via \Rp\ ({\it i.e.} 
processes corresponding to the first line of the $e^{\pm}M\!J$ and $\nu M\!J$ rows in 
table~\ref{tab:sqtopo1}), the process $\tilde{q} \rightarrow q \chi^0_1$ is simulated for 
$\chi^0_1$ masses ranging between $30\,\GeV$  and $M_{\tilde{q}}$.
In order to study the cascade gauge decays which involve two gauginos, the processes 
$\tilde{q} \rightarrow q \chi^+_1 \rightarrow q \chi^0_1 f \bar{f'}$ and
$\tilde{q} \rightarrow q \chi^0_2 \rightarrow q \chi^0_1 f \bar{f'}$ are simulated for 
$\chi^+_1$  and $\chi^0_2$ masses ranging between $40\,\GeV$ and $M_{\tilde{q}}$, and 
for $\chi^0_1$ masses between $30\,\GeV$ and $M_{\chi^+_1}$ or $M_{\chi^0_2}$. 
The masses of the $\chi$'s are varied in steps of typically $10\,\GeV$.
The lower mass values for squarks and $\chi$'s are motivated by the exclusion domains 
resulting from \Rp\ SUSY searches at LEP~\cite{LEP,L3RPV}.
The simulations allow the determination of signal detection efficiencies as a function 
of the masses of the SUSY particles involved, since the mass intervals are sufficiently 
small for linear interpolations to be used.

\section{Searches for SUSY Signals}
\subsection{Basic event selection}
The recording of the events used in this analysis is triggered using the LAr 
system~\cite{H1LARCAL}, with an efficiency close to $100\,\%$.
Background events not related to $ep$ collisions are suppressed by requiring that 
a primary interaction vertex be reconstructed within $\pm35\,\mbox{cm}$ in $z$ of 
the nominal vertex position and by using topological filters against cosmic and 
proton-beam related background.
The event time as determined by the central drift chambers is required to be consistent 
with the bunch crossing time.

\subsection{Particle identification and kinematic reconstruction}
The following criteria are used to select events containing leptons, high transverse 
momentum jets or missing transverse energy. 
An {\bf electron} is identified as an isolated and compact electromagnetic cluster 
of energy greater than $11\,\GeV$ in the LAr.
For electrons in the central detector region ($30^{\circ}<\theta_e<145^{\circ}$) a 
charged track pointing to the electromagnetic cluster is required.
A {\bf muon} candidate is identified as a track measured  in the central or forward 
tracking system, which matches geometrically with a track in the instrumented iron, 
a track in the forward muon detector or an energy deposit in the LAr calorimeter 
compatible with that expected from a minimum ionising particle.
{\bf Hadronic jets} are reconstructed from energy deposits in the calorimeter using 
a cone algorithm in the laboratory frame with a radius 
$\sqrt{\Delta \eta^2 + \Delta \phi^2} = 1$, where $\eta = -\ln \tan \frac{\theta}{2}$ 
is the pseudorapidity and $\phi$ denotes the azimuthal angle.
The {\bf missing transverse momentum} $p_{T,\rm miss}$ is obtained by the summation 
of all the energy deposits in the calorimeter.

For further selection the following Lorentz invariants are important:
  $$ y_e = 1 - \frac{E_e (1 - \cos \theta_e) }{2E_e^0}; \;\;\;\;
     Q^2_e = \frac{p^2_{T,e}}{1-y_e}; \;\;\;\;
     x_e = \frac{Q^2_e} {y_e s}; \;\;\;\;
     M_e = \sqrt{x_e s} \;\; . \;\;$$
They are determined using the measurement of the polar angle $\theta_e$, the energy 
$E_e$ and the transverse momentum $p_{T,e}$ of the electron with the highest 
$p_T$ found in the event.
$E_e^0$ denotes the energy of the incident electron.
Similar quantities can be calculated using the Jacquet-Blondel method~\cite{BLONDEL}:
 $$ y_h=\frac{\sum \left(E-p_z\right)_h}{2E_e^0}; \;\;\;\;
    Q^2_h= \frac{p^2_{T,h}}{1-y_h};\;\;\;\;
    x_h = \frac{Q^2_h} {y_h s}; \;\;\;\;
    M_h = \sqrt{x_h s} \;\; ; \;\;$$
where $p_{T,h}$ and $\sum (E-p_z)_{h}$ are calculated from the hadronic energy deposits 
in the calorimeter.

\subsection{Systematic uncertainties}
\label{sec:sys}
In each selection channel the systematic errors on the SM background expectation are 
evaluated by considering the following uncertainties.
\begin{itemize}
\item The uncertainty on the electromagnetic energy scale of the calorimeter varies 
  from $0.7\,\%$ to $3\,\%$ depending on the calorimeter region~\cite{NCCC00}.
\item The uncertainty on the hadronic energy scale is $2\,\%$.
\item The uncertainty on the integrated luminosity is $1.5\,\%$.
\item An uncertainty of $\pm7\,\%$ on the DIS expectation arises from the parton 
  densities of the proton at high $x$.
\item An uncertainty of $\pm10\,\%$ on the predicted cross section for multi-jet 
  final states is estimated by comparing the LO MC simulations where higher order QCD 
  radiation is modelled by either the CDM or DGLAP parton showers.
\end{itemize}
Furthermore, the following uncertainties related to the modelling of 
the SUSY signal are 
taken into account.
\begin{itemize}
\item The theoretical uncertainty on the signal cross 
  sections due to uncertainties in the parton densities 
varies from $7\,\%$ for $e^-u\rightarrow\tilde{d}^k_R$ at low squark masses 
  up to $50\,\%$ for $e^+d\rightarrow\tilde{u}^j_L$ at high masses.
\item Choosing either $Q^2$ or the square of the transverse momentum of the final state 
  lepton in squark decays (proceeding directly via the coupling $\lambda'$) instead of 
  $M_{\tilde{q}}^2$ as the hard scale at which the parton distributions are determined 
  yields an uncertainty of $\pm7\,\%$ on the signal cross section.  
\item An uncertainty of $10\,\%$ is attributed to the signal detection efficiencies, 
  resulting mainly from the interpolation between the simulated mass values.
\end{itemize}

\subsection{R-parity violating squark decays}
\subsubsection{Selection channel \boldmath$eq$}
The final state with an electron and a jet of high transverse momentum, resulting from 
squarks decaying in the channel $eq$, corresponds exactly to the NC DIS signature at high $x$.
However, the $M_e$ and $y_e$ distributions of the two processes differ.
Squark decays via \Rp\ lead to a resonance in the $M_e$ distribution which is measured 
with a resolution of between $3$ and $6\,\GeV$ depending on the squark mass.
Squarks produced in the $s$-channel decay isotropically in their rest frame, leading 
to a flat $d\sigma/dy$ distribution. 
In contrast, the distribution for NC DIS varies approximately as $d\sigma/dy\propto y^{-2}$.
 
The selection criteria for the $eq$ channel are the following.
\begin{itemize}
\item The total transverse momentum of the events must be balanced: $p_{T,\rm miss}<15\,\GeV$. 
\item The reconstructed momentum loss in the direction of the momentum of the incoming electron 
  must be such that $40\le\sum(E-p_z)\le 70\,\GeV$, where the sum extends over 
  all reconstructed particles. 
\item An electron must be found in the LAr calorimeter with $p_{T,e}>16\,\GeV$. 
\item To improve the sensitivity, the differences in the $M_e$ and $y_e$ distributions 
  of the SUSY signal and the DIS background are exploited by applying a lower $y_e$-cut which 
  depends on the mass of the squark under consideration. The $y_e$-cut is optimised by 
  minimising the expected limit.
  It ranges from 0.5 for masses around $100\,\GeV$ to 0.2 around $290\,\GeV$~\cite{MYTHESIS}. 
\item The selection is restricted to the kinematic range $Q_e^2>2500\,\GeV^2$ and $y_e<0.9$.
  Excluding the highest $y_e$ values avoids the region where migration effects due to initial 
  state QED radiation are largest.
  Furthermore, background from photoproduction events, in which 
hadrons are misidentified 
  as electrons, is suppressed.
\item To ensure that the various selections are exclusive\footnote{This is necessary for the 
  limit calculations (section~\ref{sec:limideri}).}, all events accepted in one of the 
  selection channels with an electron and several jets (section~\ref{sec:emjsel}) are not 
  accepted in the $eq$ channel. About 10\,\% of the candidate events are removed from the 
  $eq$ channel by this requirement.
\end{itemize}

\begin{figure}[t]
  \begin{center}
    \epsfig{file=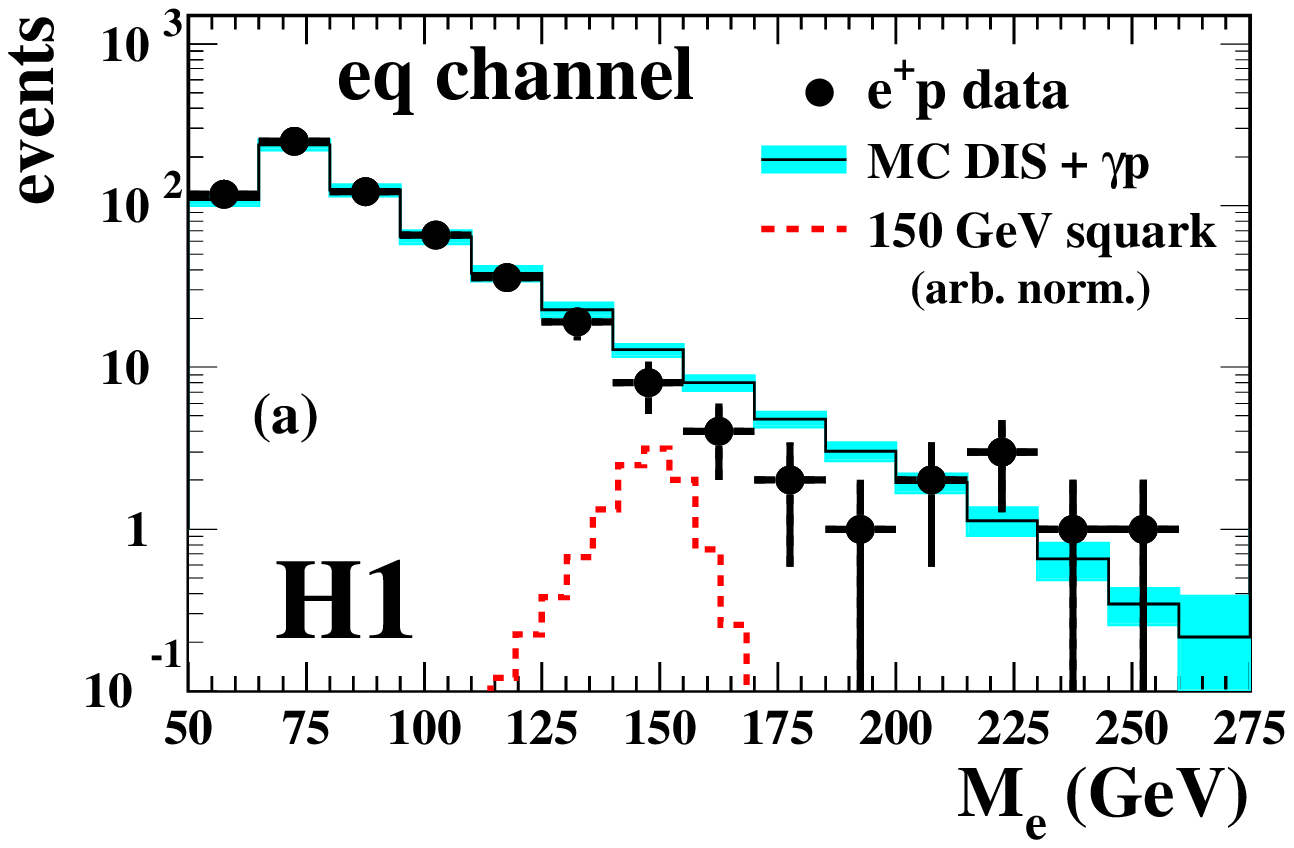, scale=0.6, angle=0.0}
    \epsfig{file=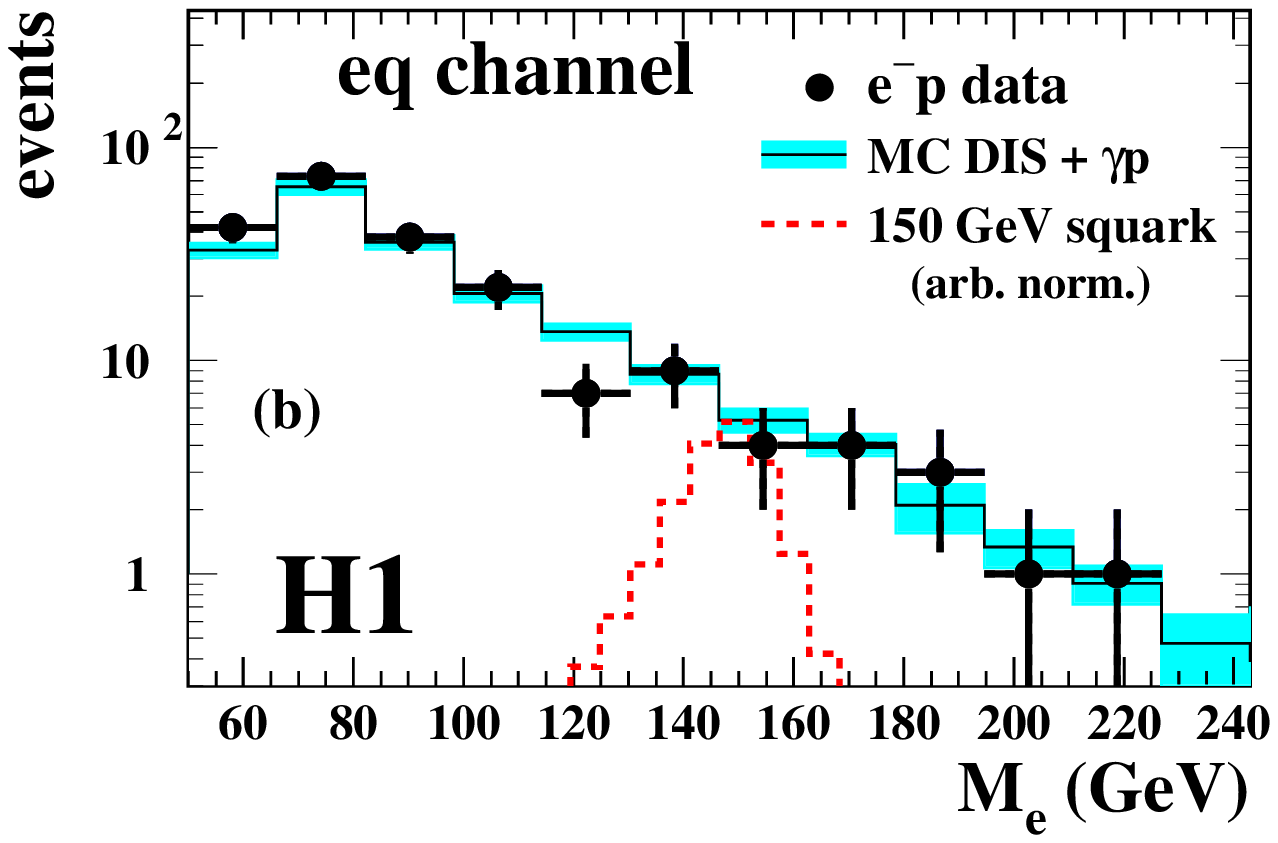, scale=0.6, angle=0.0}
  \end{center}
    \caption{ Mass spectra for the $eq$ selection channel in 
(a) $e^+p$ and (b) $e^-p$ collisions. The shaded error band indicates the 
uncertainty on the SM background. The signal expected for a squark of mass 
$150\,\GeV$ is shown with arbitrary normalisation (dashed histogram).
Events with an electron and multiple jets are not included in the spectra.}
  \label{fig:eqspectra}

\end{figure}

The $M_e$ spectra for the
data and the SM background simulation after this selection are 
shown in figure~\ref{fig:eqspectra} for $e^+p$ and $e^-p$ collisions. 
No significant deviation from the SM expectation is found for either data sample.
Table~\ref{tab:totnum} gives the total numbers of selected events and the SM expectation.
In the $e^+p$ data set $632$ candidate events are found, which is to be compared with 
$628\pm46$ expected from SM processes. 
In the $e^-p$ data sample, $204$ events are observed while the SM expectation is $192\pm14$.

\subsubsection{Selection channel \boldmath$\nu q$}

Squarks undergoing a direct \Rp\ decay into $\nu q$ lead to CC DIS-like events with high 
missing transverse momentum. The events are expected to cluster in the $M_h$ distribution 
with a resolution of $10$ to $20\,\GeV$, depending on the squark mass.

\begin{figure}[t]
  \begin{center}
    \epsfig{file=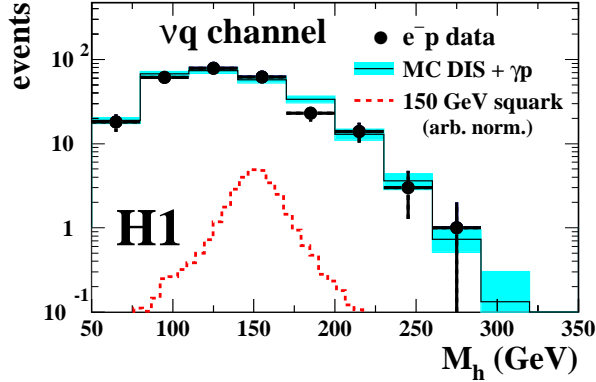, scale=0.6, angle=0.0}
  \end{center}
    \caption{Mass spectrum for the $\nu q$ selection channel in $e^-p$ collisions.
  The shaded error band indicates the uncertainty on the SM background.
  The signal expected for a squark of mass $150\,\GeV$ is shown 
with arbitrary normalisation (dashed histogram).
  Events accepted in the selection channel $\nu MJ$ are not included in the spectra.}
\label{fig:nqspectrum}

\end{figure}

The selection criteria for the $\nu q$ channel are the following:
\begin{itemize}
\item The missing transverse momentum must be greater than $30\,\GeV$. 
\item No electron or muon must be found with $p_{T}>5\,\GeV$.
\item The events must lie in the kinematic range $Q_h^2>2500\,\GeV^2$ and $y_h<0.9$. 
  The resolutions in both $M_h$ and $Q_h^2$ degrades with increasing $y_h$ since 
  both $\delta M_h/M_h$ and $\delta Q^2_h/Q^2_h$ behave as $1/(1-y_h)$ for $y_h\sim 1$. 
  Hence the high $y_h$ range is excluded.
\item To ensure exclusivity with respect to the $\nu MJ$ channel (section~\ref{sec:numj}), 
  events with two or more jets with $p_{T,\rm jet}>15\,\GeV$ are rejected.
  This removes about $3.5\,\%$ of the candidate events.
\end{itemize}
Only $\tilde{d}_R$-type squarks, which are produced mainly in $e^-p$ collisions, can undergo 
a decay into the $\nu q$ final state. The $M_h$ spectrum of this data set and the SM background 
are shown in figure~\ref{fig:nqspectrum}. No significant deviation from the SM expectation is found.
$261$ events are observed in the data
and $269\pm21$ are expected according to the SM.

\subsection{Squark gauge decays leading to {\boldmath{$e$}} + jets + \boldmath{$X$} final states}
\label{sec:emjsel}
For the channels $e^+MJ$, $e^-MJ$, $ee MJ$, $e\mu MJ$ and $\nu eMJ$ a common preselection 
is carried out:
\begin{itemize}
\item At least one electron must be found with $p_{T,e}>6\,\GeV$ in the angular 
range $5^{\circ}<\theta_e<110^{\circ}$.  For central electrons ($\theta_e>30^{\circ}$) the 
charged track, measured in the central tracking system, must geometrically and kinematically 
match the electromagnetic cluster. To discriminate against fake-electron background from 
photoproduction, electron candidates in the forward region ($\theta_e<30^{\circ}$) have to 
fulfill harsher isolation criteria and 
the $\sum (E-p_z)$ of the event must be greater than $30\,\GeV$. 
The latter cut causes only a small efficiency loss for all channels discussed here.
\item At least two jets must be found with $p_{T,\rm jet}> 15\,\GeV$ in the range 
  $7^{\circ} < \theta_{\rm jet}< 145^{\circ}$.
\item For all final state topologies considered here, the squark decay products are mainly 
  emitted in the forward direction. This is exploited by requiring that: 
\begin{itemize}
\item $Q^2_e>1000\,\GeV^2$.
\item At least one of the polar angles of the highest $p_T$ electron and 
the two 
  highest $p_T$ jets is less than $40^\circ$.
\item Of the two jets with highest $p_T$, that with the larger polar angle $\theta_{\rm backw}$ 
  satisfies $\theta_{\rm backw}<180^{\circ} \cdot(y_e-0.3)$. This cut efficiently separates 
  the SUSY signal events from the NC DIS background~\cite{MYTHESIS}. 
\end{itemize}
\end{itemize}
After this preselection, 91 (22) events are found
in the $e^+p$ ($e^-p$) data sample while 
$89.3\pm3.7$ ($22.6\pm0.7$) is the SM expectation. Further cuts are applied for each sub-channel.

\begin{figure}[t]
  \begin{center}
    \epsfig{file=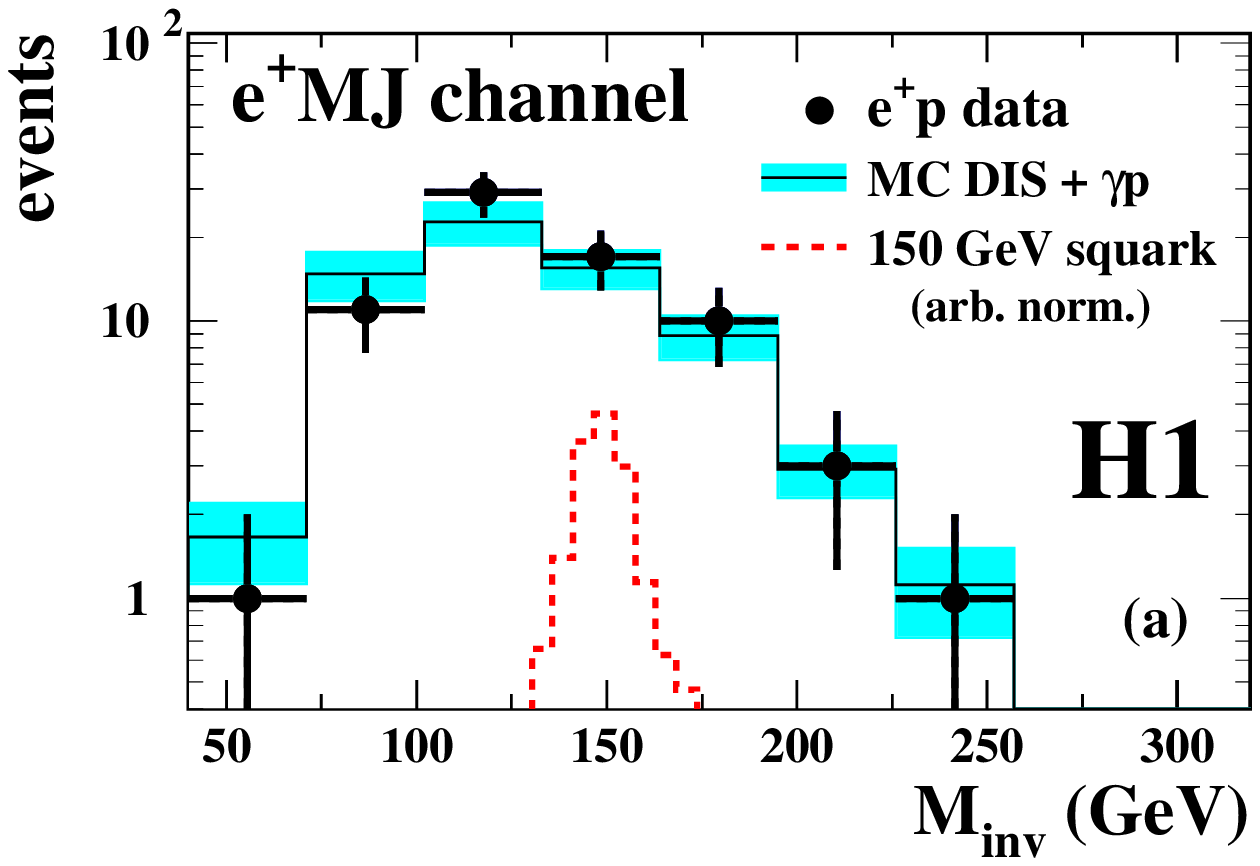, scale=0.6, angle=0.0}
    \epsfig{file=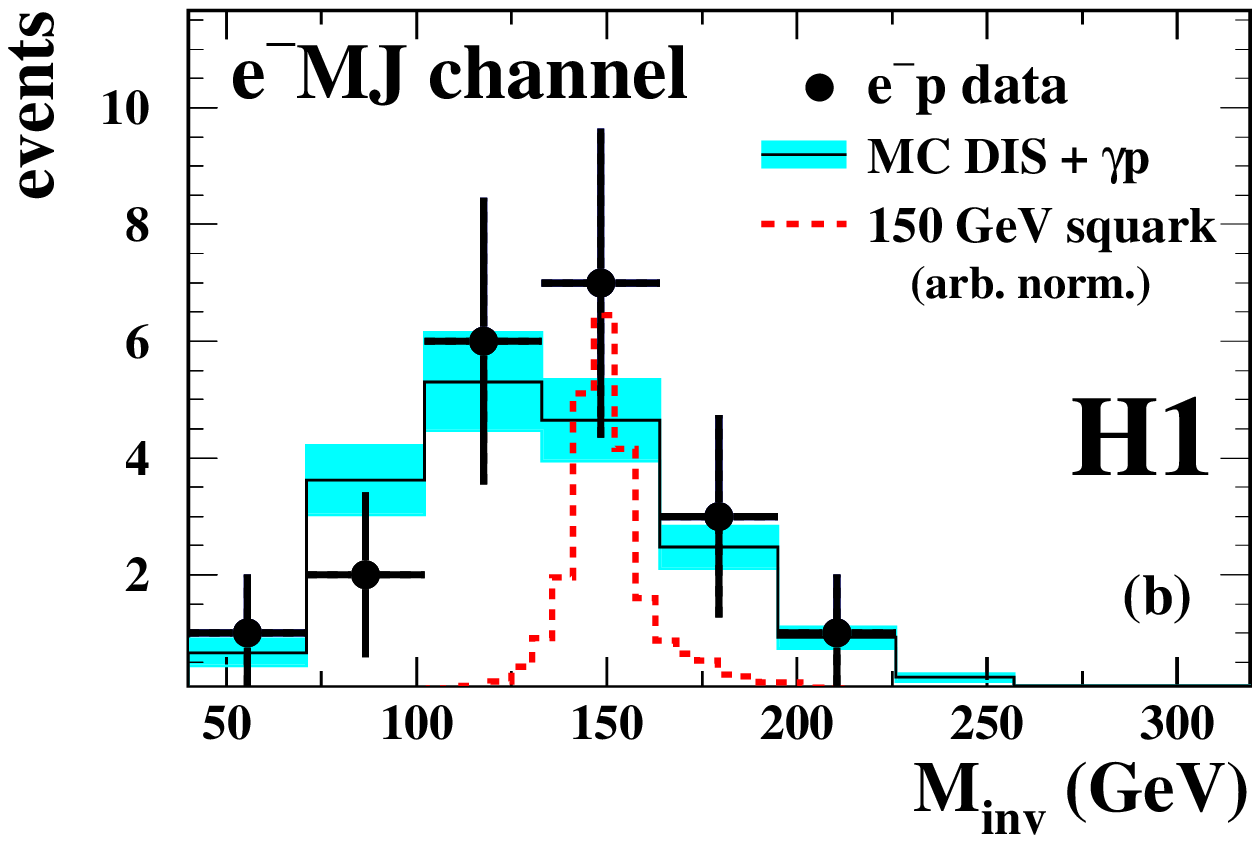, scale=0.6, angle=0.0}
  \end{center}
    \caption{ Mass spectra for (a) the $e^+MJ$ selection channel 
in $e^+p$ collisions  and (b) the $e^-MJ$ selection channel in $e^-p$ collisions.
The shaded error band indicates the uncertainty on the SM background.
The signal expected for a squark of mass $150\,\GeV$ is shown 
with arbitrary normalisation (dashed histogram).}
  \label{fig:emjspectra}
\end{figure}

\subsubsection{Channels with ``wrong'' and ``right'' lepton charge}
For the channels $e^+MJ$ and $e^-MJ$ no neutrinos are involved in the final state.
Therefore the missing momentum 
is restricted according to $p_{T,\rm miss}<15\,\GeV$ 
and $40<\sum(E-p_z)<70\,\GeV$. To ensure that the selection is exclusive with respect 
to the $eeMJ$ and $e\mu MJ$ channels, events with a second electron with $p_{T,e}>5\,\GeV$ 
and $5^{\circ}< \theta_e< 110^{\circ}$, or a muon with $p_{T,\mu}>5\,\GeV$ and 
$10^{\circ}< \theta_{\mu}< 110^{\circ}$, are rejected.

Events are accepted in the channel having the ``wrong'' charge lepton, {\it i.e.} different 
from the incident beam, if the electron/positron is found in the angular range 
$\theta_e>30^{\circ}$ (where the charge measurement is made with the central tracking system) 
and the charge is measured to be opposite to that of the incident lepton, with a significance 
greater than two standard deviations.
No candidates are found in the data and the SM expectation in this channel is very low 
(see table~\ref{tab:totnum}).

In the ``right'' charge lepton channel, {\it i.e.} the same charge as the lepton beam, events 
are accepted if they contain either a central electron ($\theta_e>30^{\circ}$) with a charge 
measurement of the ``right'' sign or an electron found in the forward region ($\theta_e<30^{\circ}$).
In the latter case no charge requirement is made.
For the selected events, an invariant squark mass $M_{\rm inv}$ is calculated as:
$M_{\rm inv}=\sqrt{4 E^0_e \left( \sum_{i} E_i - E^0_e \right) }$,
where the sum runs over the electrons and the jets found in the event with $p_{T}>5\,\GeV$.
This method yields a good reconstruction of the squark mass with a typical resolution of 
$7$ to $10\,\GeV$.
The  $M_{\rm inv}$ distributions for the data and 
the SM expectation are shown 
for the ``right'' charge $eMJ$ channel 
in figure~\ref{fig:emjspectra} for $e^+p$ and $e^-p$ collisions.
No significant deviation from the SM is observed at any mass value.
In total 72 (20) events are selected in the $e^+p$ ($e^-p$) data set with 
$67.5\pm 9.5$ ($17.9\pm 2.4$) predicted from SM background processes.

\subsubsection{Channels with an additional lepton}
The further selection for the channels $eeMJ$ and $e\mu MJ$ requires either an additional 
electron with the same criteria as described in the common preselection,  or an additional 
muon with $p_{T,\mu}>5\,\GeV$ in the polar angle range $10^{\circ}<\theta_{\mu}<110^{\circ}$.
To ensure exclusivity, events are accepted in one selection channel ($eeMJ$ or $e\mu MJ$) 
only\footnote{Events with a muon with $p_{T,\mu}>5\,\GeV$ and 
$10^{\circ}<\theta_{\mu}<110^{\circ}$ are not accepted in the $eeMJ$ channel. Similarly, events 
with an additional electron in the range $5^{\circ}<\theta_{e}<110^{\circ}$ are not accepted 
in the $e\mu MJ$ channel.}.
In these channels the SM background (mainly NC DIS) is very low (see table~\ref{tab:totnum}).
No candidate events for either $eeMJ$ or $e\mu MJ$ are found in either data set, which is 
compatible with the SM expectation.

For the channel $\nu eMJ$, candidate events, possibly containing a neutrino, are selected 
by requiring $p_{T,\rm miss}>15\,\GeV$. A cut of $y_e(y_e-y_h)>0.04$ exploits the fact that 
for the SUSY signal the escaping neutrino carries  a non-negligible part of 
$\sum \left( E-p_z \right) $ and hence the variable $y_h$ is substantially smaller than $y_e$,
while $y_e \sim y_h$ is expected for NC DIS events.
Events previously accepted in the $eeMJ$ or $e\mu MJ$ channels are rejected.
No events are found in this channel.
This is compatible with the SM expectation (mainly NC DIS) as detailed in table~\ref{tab:totnum}.

\subsection{Squark gauge decays leading to {\boldmath$\nu$} + jets + \boldmath{$X$} final states}
\label{sec:numj}
\begin{figure}[t]
  \begin{center}
    \epsfig{file=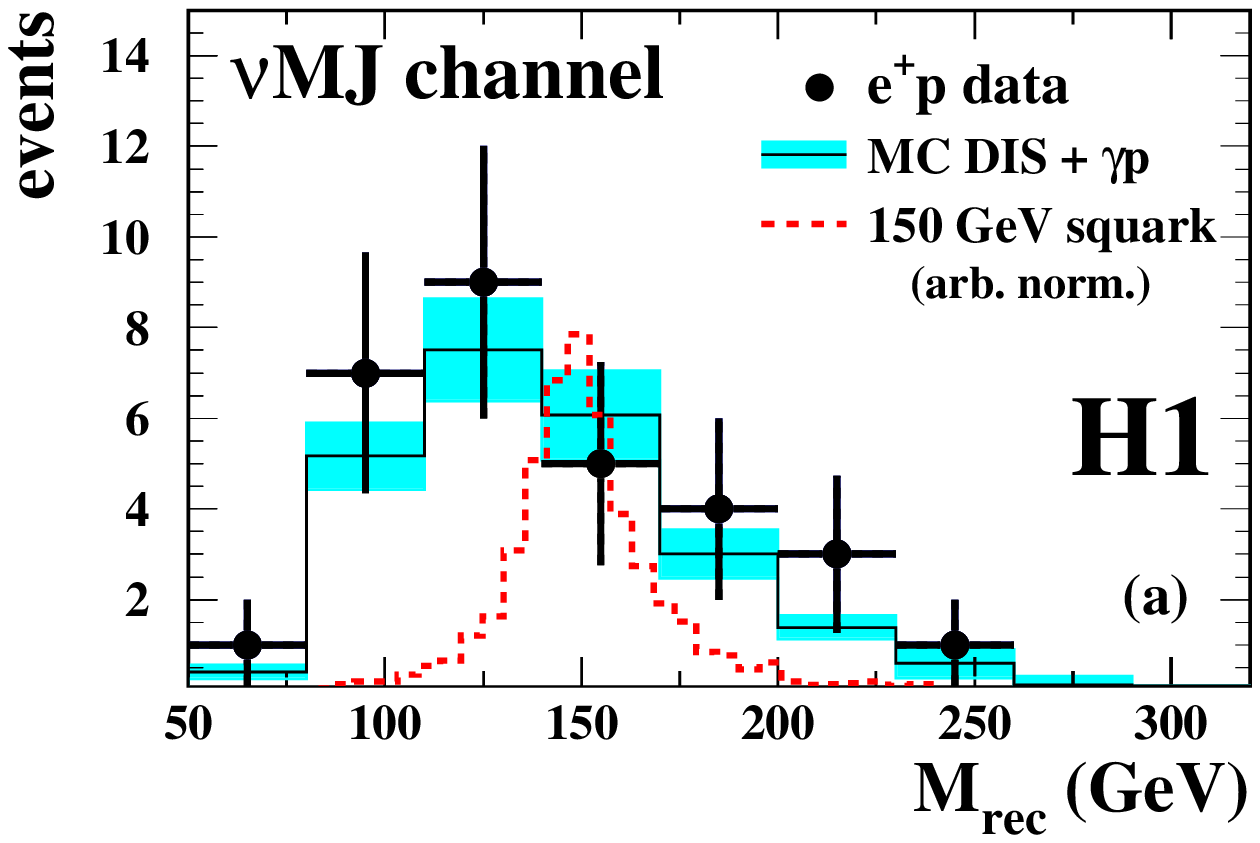, scale=0.6, angle=0.0}
    \epsfig{file=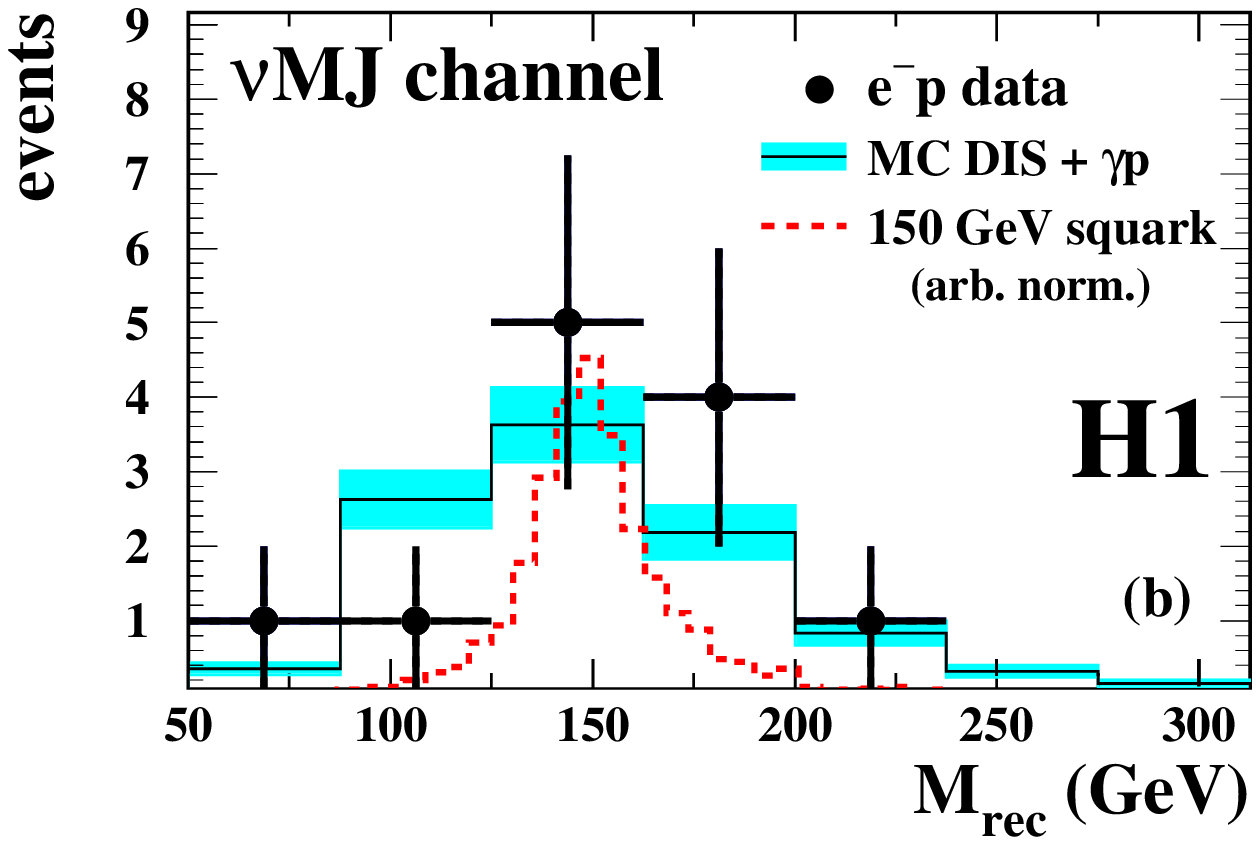, scale=0.6, angle=0.0}
  \end{center}
    \caption{Mass spectra for the $\nu MJ$ selection channel in (a) $e^+p$ and (b) $e^-p$ collisions.
The shaded error band indicates the uncertainty on the SM background.
The signal expected for a squark of mass $150\,\GeV$ is shown 
with arbitrary normalisation (dashed histogram).}
\label{fig:numjspectra}
\end{figure}
The selection of $\nu MJ$ and $\nu\mu MJ$ candidates starts with the requirement that:
\begin{itemize}
\item The missing transverse momentum satisfies $p_{T,\rm miss}>26\,\GeV$.
\item At least two jets with $p_{T,\rm jet}> 15\,\GeV$ are reconstructed in the angular 
range $7^{\circ} < \theta_{\rm jet}< 145^{\circ}$.
\item No electron is found in the event.
\end{itemize}
If no muon is found, the event is identified as a $\nu MJ$ candidate.
Assuming that the missing energy of a candidate event is carried by one neutrino only, 
its kinematics are reconstructed by exploiting energy-momentum conservation.
The four-vector of this $\nu$ is then added to that of the hadronic final state to reconstruct 
the invariant mass $M_{\rm rec}$ of the incoming electron and quark.
The mass resolution of this method is about $15\,\GeV$.
The $M_{\rm rec}$ spectra of the data and the expected SM background are shown in 
figure~\ref{fig:numjspectra}.
In the $e^+p$ ($e^-p$) data set 30 (12) $\nu MJ$ candidate events are selected while 
$24.3\pm3.6$ ($10.1\pm 1.4$) are expected from SM background (mainly CC DIS). 

If in addition to the above requirements a muon with $p_{T,\mu}>5\,\GeV$ in the 
range $10^{\circ}<\theta_{\mu}<110^{\circ}$ is found, the events are identified 
as $\nu\mu MJ$ candidates. No candidate events are found in either data set.
This is compatible with the SM expectation (predominantly CC DIS), which is shown 
in table~\ref{tab:totnum}.

\section{Exclusion Limits}
\label{sec:limideri}
\begin{table}[t]
 \begin{center}
  \begin{tabular}{|c|cc|cc|c|}
  \hline
  \rule[-2mm]{0mm}{7mm} & \multicolumn{2}{c|}{ \bf \boldmath $e^+p$ collisions} & \multicolumn{2}{c|}{ \bf \boldmath $e^-p$ collisions} &\\  
\multicolumn{1}{|c|}{\bf Channel} & \multicolumn{1}{c}{\bf Data} & {\bf SM expectation} & {\bf Data} & {\bf SM expectation} & {\bf Efficiency}\\ \hline \hline
  {{\boldmath$eq$}} & 632 & 628\,$\pm$\,46 & 204 & 192\,$\pm$\,14 &$30-50\,\%$\\
  {{\boldmath$\nu q$}} & ---& --- & 261  & 269\,$\pm$\,21 &$40-60\,\%$\\\hline
  {{\boldmath\bf$eMJ$ (``right'' charge)}} & 72 & 67.5\,$\pm$\,9.5  & 20 & 17.9\,$\pm$\,2.4 &$15-50\,\%$\\
  {{\boldmath\bf$eMJ$ (``wrong'' charge) }} & 0 & 0.20\,$\pm$\,0.14 & 0 & 0.06\,$\pm$\,0.02 &$10-30\,\%$\\
  {{\boldmath$ee MJ$}} & 0 & 0.91\,$\pm$\,0.51   & 0 & 0.13\,$\pm$\,0.03 &$15-45\,\%$\\
  {{\boldmath$e\mu MJ$}} & 0 & 0.91\,$\pm$\,0.38 & 0 & 0.20\,$\pm$\,0.04 &$15-35\,\%$\\
  {{\boldmath$\nu eMJ$}} & 0 & 0.74\,$\pm$\,0.26& 0 & 0.21\,$\pm$\,0.07 &$15-40\,\%$\\\hline
  {{\boldmath$\nu MJ$}} & 30 & 24.3\,$\pm$\,3.6 & 12 & 10.1\,$\pm$\,1.4 & $10-60\,\%$\\
  {{\boldmath$\nu\mu MJ$}} & 0 & 0.61\,$\pm$\,0.12& 0 & 0.16\,$\pm$\,0.03 &$15-50\,\%$\\\hline
 \end{tabular}
 \end{center}
  \caption[Selection summary: total event numbers] {Total numbers of selected events, SM 
expectations and ranges of selection efficiencies of the squark decay channels considered 
in $e^+p$ and in $e^-p$ collisions. The $\tilde{u}_L$-type squarks ($e^+p$ collisions) cannot 
decay to $\nu q$.}
  \label{tab:totnum}
\end{table}

The total numbers of selected and expected events are summarised in table~\ref{tab:totnum} 
for all final state topologies considered in this analysis. It is assured by the choice of 
the selection cuts for the different channels that the selection of all topologies is 
fully exclusive. No significant deviation from the SM expectation is found in any channel.
The selection channels are combined, separately for the $e^+p$ and $e^-p$ data sets, to 
derive constraints on \Rp\ SUSY models.
\subsection{Method of limit derivation}
For a given set of parameters in a certain supersymmetric model, the full supersymmetric 
mass spectrum and the branching ratios of all squark decay modes are calculated using 
the SUSYGEN package.
An upper limit $N_{\rm lim}$ on the number of events coming from squark production 
is calculated at a confidence level (CL) of $95\,\%$ using a modified frequentist 
approach based on Likelihood Ratios~\cite{JUNK}.
The following quantities enter the limit calculation.
\begin{itemize}
\item The numbers of events observed in the data for all selection channels. For the channels 
where the SM background is considerable ($eq$, $\nu q$, $e MJ$ with ``right'' lepton charge 
and $\nu M\!J$)  the event numbers are integrated within a mass bin around the squark mass 
under consideration. 
For each decay channel the width of the mass bin is optimised, {\it i.e.} the expected limit 
is minimised, using the reconstructed mass distributions from the SUSY signal and SM background simulations. 
For the channels  $eMJ$ with ``wrong'' lepton charge, $eeMJ$, $e\mu MJ$, 
$\nu eMJ$  and $\nu \mu M\!J$, no mass restriction is imposed, since the SM backgrounds 
are small. 
\item The event numbers expected from SM background processes and their systematic uncertainties.
\item The signal detection efficiencies (see table~\ref{tab:totnum}) and their uncertainties 
for all squark decay processes in all selection channels, 
obtained using the calculated spectrum 
of sparticle masses.
\item The calculated branching ratios of all squark decay modes. 
\end{itemize}
A bound on the squark production cross section $\sigma_{\rm lim}$ is then obtained 
from $N_{\rm lim}$. Sets of model parameters that lead to signal cross sections 
above $\sigma_{\rm lim}$ can be excluded.

The case of non-vanishing Yukawa couplings $\lambda'_{131}$ or $\lambda'_{113}$, which 
correspond to the resonant production of stop and sbottom squarks, is treated 
separately since the top and bottom quark masses cannot be neglected in the calculation 
of couplings and branching ratios. Furthermore, a top quark could be produced in gauge decays.
The top quark decays via $t \rightarrow b W$, leading to decay products different from those 
of the first two generations for which the efficiencies are determined. 
Diagrams 
which lead to a top in the final state are thus not 
taken into account in the calculation of 
the branching ratios. This represents a conservative approach, 
since most of the top decays are implicitly 
covered in the 
selection channels and would be visible in the mass distributions and the 
total event numbers.

\subsection{Limits in the ``phenomenological'' MSSM}
\begin{figure}[p]
  \begin{center}
    \epsfig{file=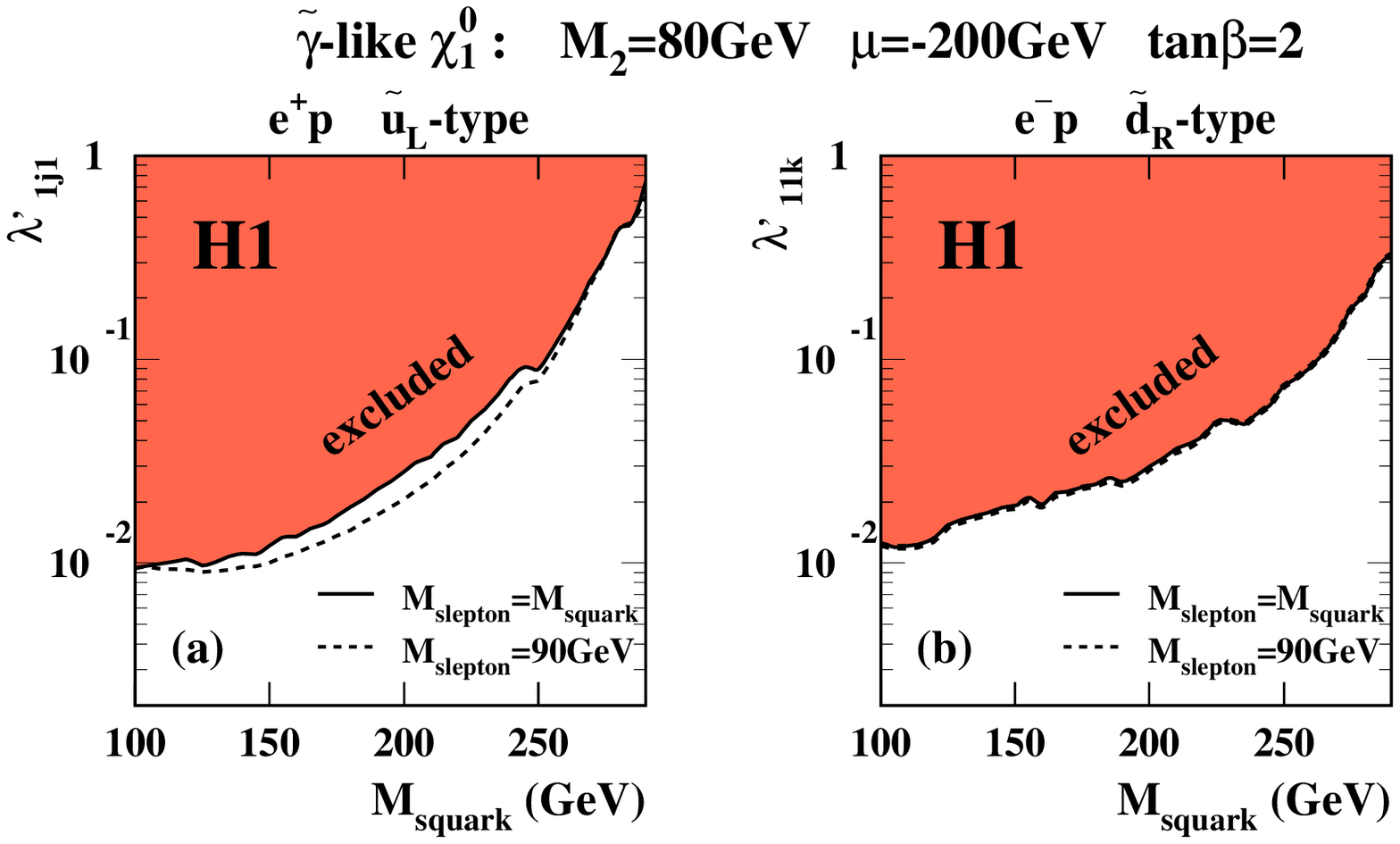, scale=0.7, angle=0.0}
    \epsfig{file=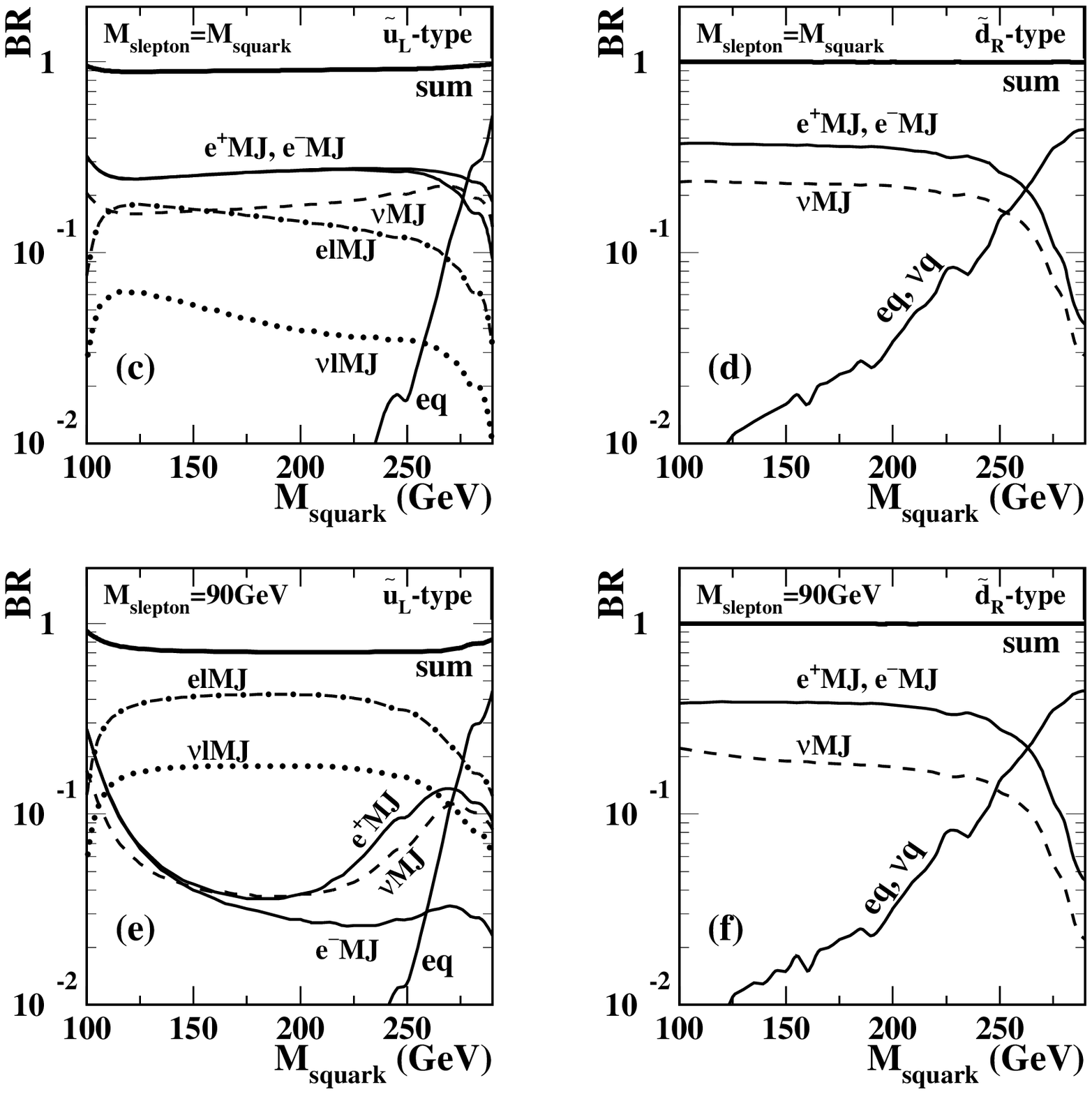, scale=0.7, angle=0.0}
  \end{center}
    \caption{
(a, b) Exclusion limits at the $95\,\%$ CL on (a) $\lambda'_{1j1}$ with $j=1,2$ and (b) 
$\lambda'_{11k}$ with $k=1,2$. (c -- f) Branching ratios to the decay channels 
considered in the analysis for $\lambda'$ values at the exclusion limits shown in 
(a) and (b). The results are shown for MSSM parameters leading to a $\chi_1^0$ 
dominated by its photino component when slepton and squark masses are assumed to be 
degenerate (c, d) and for a slepton mass of $90\,\GeV$ (e, f).}
   \label{fig:scena}
\end{figure}

A version of the MSSM is considered here
where the masses of the neutralinos, charginos 
and gluinos, as well as the couplings between any two SUSY particles and a SM 
fermion/boson, are determined by the usual parameters.
These are the ``mass'' term $\mu$, which mixes the Higgs superfields, the SUSY soft-breaking
mass parameters $M_1$, $M_2$ and $M_3$ for $U(1)$, $SU(2)$ and $SU(3)$ gauginos, respectively,
and the ratio $\tan \beta$ of the vacuum expectation values of the two neutral scalar Higgs fields.
The parameters are defined at the electroweak scale. 
The gaugino mass terms are assumed to
unify at a Grand Unification (GUT) scale to a
common value $m_{1/2}$, leading to the usual relations~\cite{MSSM} between $M_1$, $M_2$ and $M_3$.
The gluino mass is approximated by the value of $M_3$ at the electroweak scale.
The sfermion masses are free parameters in this model. Possible mixing 
between sfermions is neglected and all squarks are assumed to
be degenerate in mass. The 
possibility of a photino-like $\chi_1^0$ 
is first discussed, before turning to a complete scan of the SUSY 
parameter space.

\subsubsection{Exclusion limits for a photino-like \boldmath{$\chi_1^0$}}
For an example set of the MSSM parameters ($\mu=-200\,\GeV$, $M_2=80\,\GeV$, $\tan\beta=2$) 
leading to a $\chi_1^0$ dominated by its photino component, exclusion limits at the 95\,\% CL 
on $\lambda'_{1j1} (j=1,2)$ and $\lambda'_{11k} (k=1,2)$ are shown in figures~\ref{fig:scena} 
(a, b) as a function of the squark mass. The full curves represent cases in which sleptons 
and squarks are assumed to be degenerate in mass. The dashed curves indicate the limits for 
slepton masses $M_{\tilde{l}}$ fixed at $90\,\GeV$, close to the lowest 
mass bound from \Rp\ sfermion searches 
at LEP~\cite{LEP}.  The HERA sensitivity allows tests of \Rp\ Yukawa couplings $\lambda'$ 
down to around $10^{-2}$ for squark masses of $100\,\GeV$. For a high squark mass the 
sensitivity degrades since the production cross section decreases. At a squark mass of 
$290\,\GeV$, $\lambda'_{1j1}$  ($\lambda'_{11k}$) values larger than 0.6 (0.3) are ruled out.

The branching ratios to all channels calculated for a $\lambda'$ value exactly at the exclusion 
limit are illustrated in figures~\ref{fig:scena} (c -- f). The total branching fraction covered 
exceeds $75\,\%$ for all points in the MSSM parameter space and is generally close to 100\,\%.
At large squark masses, a large Yukawa coupling $\lambda'$ is necessary to allow visible squark 
production. As a result the decay channels $eq$ and $\nu q$ proceeding directly via $\lambda'$ 
become important. For smaller masses, the dominant channels in the case of a photino-like 
$\chi_1^0$ are those with an $e^{\pm}$ and several jets in the final state.

For $\tilde{u}_L$-type squarks ($\lambda'_{1j1}\ne 0$) the relative contributions of the 
gauge decay channels strongly depend on the slepton mass. In the case of a light slepton 
($M_{\tilde{l}}=90\,\GeV$), the decays of a $\chi_1^+$ into a lepton-slepton pair are 
kinematically allowed. Thus cascade gauge decays of $\tilde{u}_L$-type squarks are possible, 
leading to enhanced contributions from the channels $elMJ$ and $\nu l MJ$.
In contrast, the cascade gauge decays of $\tilde{u}_L$-type squarks are kinematically 
suppressed for $M_{\tilde{l}}=M_{\tilde{q}}$. The dependence of the $\lambda'_{1j1}$ limit 
on the slepton mass is rather small since the sensitivities of all selection channels are 
similar. In the case of $\tilde{d}_R$-type squarks ($\lambda'_{11k}\ne 0$), the relative 
contributions of the decay channels and the resulting limit on $\lambda'_{11k}$ are 
almost independent of the slepton mass, since gauge decays of $\tilde{q}^{}_R$ squarks 
via charginos are suppressed.

The branching ratios to the various 
decay channels depend on the SUSY parameters.
Thus, for parameter values different from those discussed above,
different decay channels are dominant. For instance, for a zino-like 
$\chi_1^0$ the dominant channels at lower squark masses are those with 
a $\nu$ and several jets in the final state~\cite{MYTHESIS}.
Cascade decays of $\tilde{q}^{}_R$ squarks are also possible for 
some parameter configurations via gauge decays involving neutralinos or gluinos.

\subsubsection{Scan of the parameter space}
\begin{figure}[t] 
  \begin{center}
    \epsfig{file=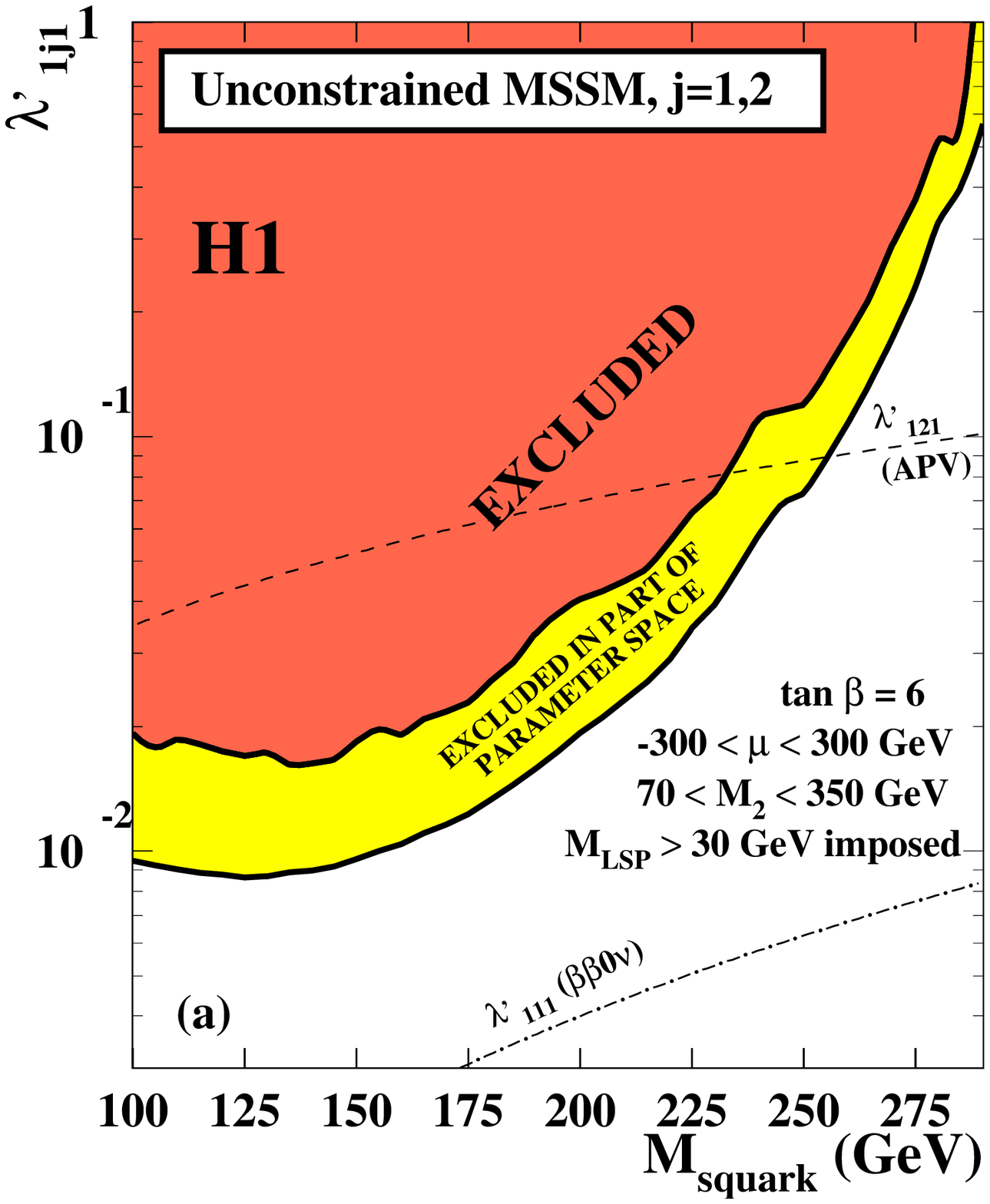,width=7.9cm}
    \epsfig{file=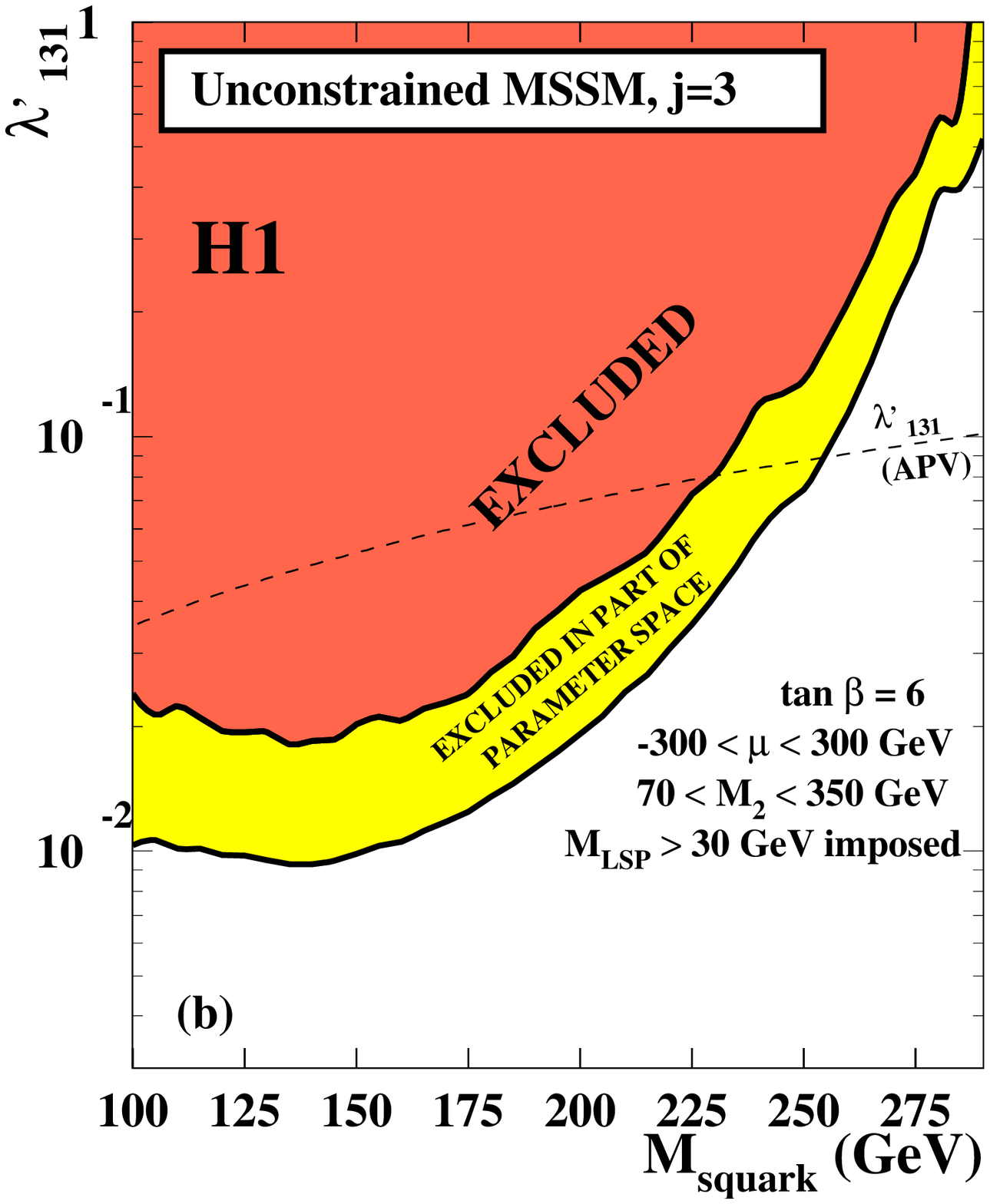,width=7.9cm}
  \end{center}
  \caption{Exclusion limits  ($95\,\%$ CL) on $\lambda'_{1j1}$ for (a) $j=1,2$ and 
(b) $j=3$ as a function of the squark mass from a scan of the MSSM parameter space as 
indicated in the figures. The two full curves indicate the strongest and the weakest 
limits on $\lambda'$ in the parameter space investigated. Indirect limits from neutrinoless 
double beta decay experiments ($\beta\beta 0\nu$) and atomic parity violation (APV) 
are also shown.}
  \label{fig:scanj}
\end{figure} 

\begin{figure}[t] 
  \begin{center}
    \epsfig{file=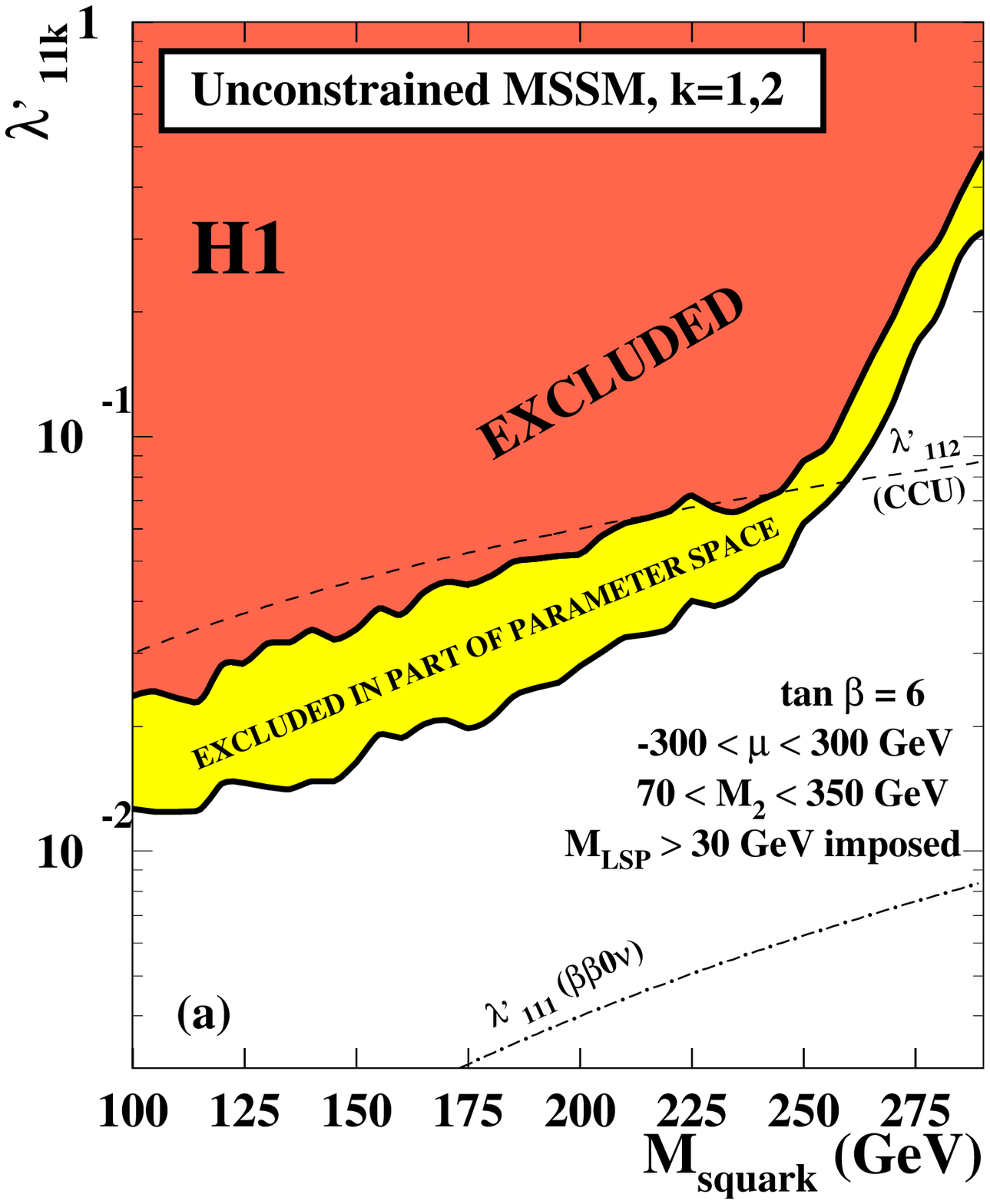,width=7.9cm}
    \epsfig{file=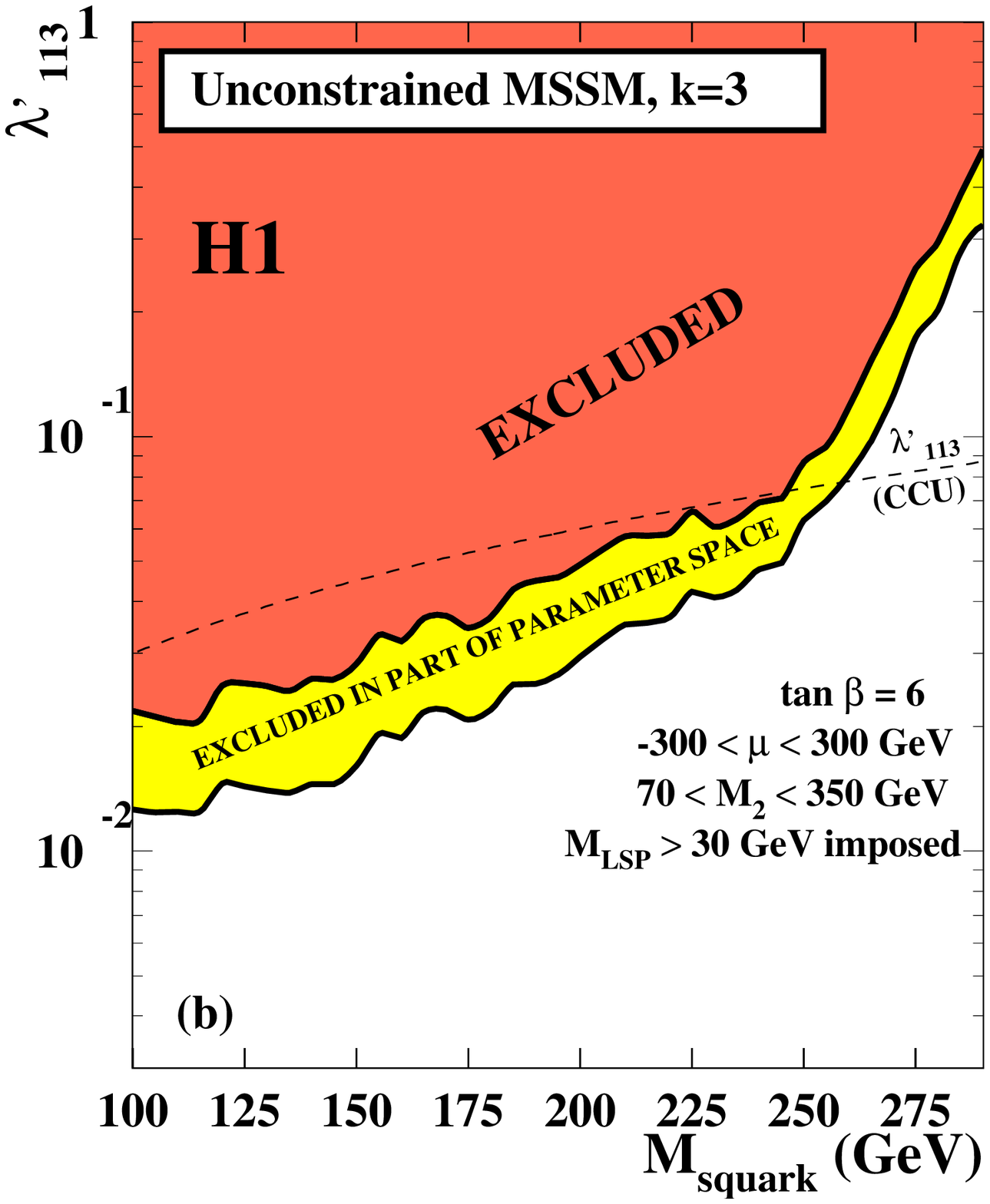,width=7.9cm}
  \end{center}
  \caption{Exclusion limits ($95\,\%$ CL) on $\lambda'_{11k}$ for (a) $k=1,2$ and (b) 
$k=3$ as a function of the squark mass from a scan of the MSSM parameter space. The two full 
curves indicate the strongest and the weakest limits on $\lambda'$. Indirect limits from 
neutrinoless double beta decay experiments ($\beta\beta 0\nu$) and 
tests of charged current 
universality (CCU) are also shown. }
  \label{fig:scank}
\end{figure} 

In order to investigate the dependence of the sensitivity on the MSSM parameters, a scan of 
$M_2$ and $\mu$ is performed for $\tan \beta = 6$.  Again, sleptons are assumed to be 
degenerate and their mass is set to a fixed value of $90\,\GeV$. Other values for 
$M_{\tilde{l}}$ and $\tan\beta$ lead to very similar results.
The parameters $M_2$ and $\mu$ are varied in the range 
$70\, {\rm GeV} < M_2 < 350\,\GeV$ 
and {\mbox{$-300\,\GeV < \mu < 300\,\GeV$}}. Parameter sets leading to a scalar LSP or to 
LSP masses below $30\,\GeV$ are not considered. The latter restriction, as well as the lower 
value for $M_2$, are motivated by the exclusion domains resulting from $\chi$ searches in 
\Rp\ SUSY at LEP~\cite{L3RPV}. Upper bounds on 
the couplings $\lambda'_{1j1}$ and $\lambda'_{11k}$ are obtained
for each point in the $(\mu, M_2)$ plane. 
The results are shown for $\lambda'_{1j1}$ in figure~\ref{fig:scanj}
and for $\lambda'_{11k}$ in figure~\ref{fig:scank}. 
For each plot, the 
two full 
curves indicate the strongest and weakest limits obtained for $\lambda'$ in 
the parameter space investigated. As can be seen from the narrowness of the region that 
is excluded in only part of the parameter space, the limits on both $\lambda'_{1j1}$ 
and $\lambda'_{11k}$ are widely independent of the SUSY parameters.
For a Yukawa coupling of electromagnetic strength, {\it i.e.} 
$\lambda'_{1j1}=\sqrt{4\pi\alpha_{\rm em}}=0.3$ ($\lambda'_{11k}= 0.3$), 
$\tilde{u}_L$, $\tilde{c}_L$ and $\tilde{t}_L$ 
($\tilde{d}_R$, $\tilde{s}_R$ and $\tilde{b}_R$) squarks with 
masses below $\sim 275\,\GeV$  ($280\,\GeV$) are excluded at the $95\%$ CL. 
For a coupling strength smaller by a factor of 100,
masses up to $\sim 220\,\GeV$ are ruled out. 

In figures~\ref{fig:scanj} and \ref{fig:scank} the results for the direct production 
of squarks are compared with indirect limits from virtual squark exchange in low 
energy experiments~\cite{INDIRDREINER}. The production of  $\tilde{u}$ and $\tilde{d}$ 
squarks via a $\lambda'_{111}$ coupling is 
tightly constrained by the non-observation 
of neutrinoless double beta decay ($\beta\beta0\nu$)~\cite{BETA0NU}.
The best indirect limit on the couplings $\lambda'_{121}$ and $\lambda'_{131}$ comes 
from atomic parity violation (APV) measurements~\cite{INDIRDREINER,APV}.
The best indirect limit on the couplings $\lambda'_{112}$ and $\lambda'_{113}$ results 
from tests of charged current universality (CCU)~\cite{CCU}.
The HERA results improve the limits on 
$\lambda'$ for squarks of the second and third family 
({\it i.e.} $\lambda'_{121}, \lambda'_{131}, \lambda'_{112}, \lambda'_{113}$) 
for masses up to $\sim255\,\GeV$.

\subsection{Limits in the minimal Supergravity model}

In this section the minimal Supergravity (mSUGRA) model~\cite{MSUGRA}
is considered,
where the number of free parameters is reduced by assuming, in addition to the GUT 
relation between $M_1$, $M_2$ and $M_3$ mentioned previously, a universal mass parameter 
$m_0$ for all scalar fields at the GUT scale. By requiring in addition Radiative 
Electroweak Symmetry Breaking (REWSB) the model is completely determined by 
$m_0$, $m_{1/2}$, $\tan \beta$, the sign of $\mu$ and the common trilinear coupling 
at the GUT scale $A_0$. The modulus of $\mu$ is related to the other model parameters.
The program SUSPECT~2.1~\cite{SUSPECT} is used to obtain the REWSB solution for 
$|\mu|$ and calculate the full supersymmetric mass spectrum.

Assuming a fixed value for the $\Rp$ couplings $\lambda'_{1j1}$ and $\lambda'_{11k}$, 
constraints on the mSUGRA parameters can be set, for example on $(m_0, m_{1/2})$, when 
$\tan \beta$, $A_0$, and the sign of $\mu$ are fixed. $A_0$ enters only marginally in 
the interpretation of physics results at the electroweak scale and it is set to zero.
The efficiencies for the detection of all gauge decays of squarks involving a gaugino 
lighter than $30\,\GeV$ are set to zero since the parameterisation of the efficiencies 
is not valid in this domain. The corresponding parameter space is already excluded by 
$\chi$ searches in \Rp\ SUSY at LEP~\cite{L3RPV}.

\subsubsection{Results for the first and second families}

\begin{figure}[t] 
  \begin{center}
    \epsfig{file=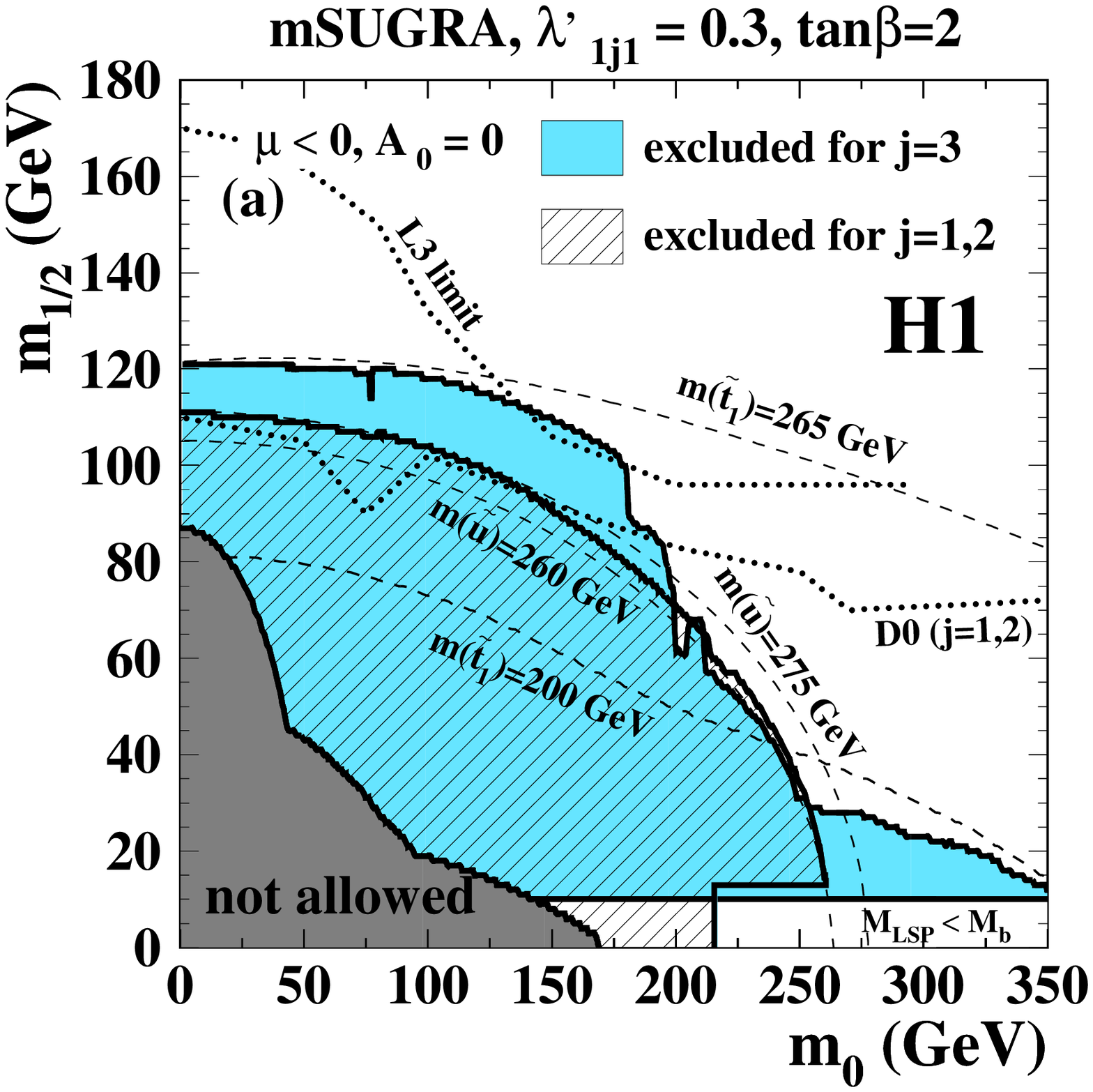,width=7.9cm}
    \epsfig{file=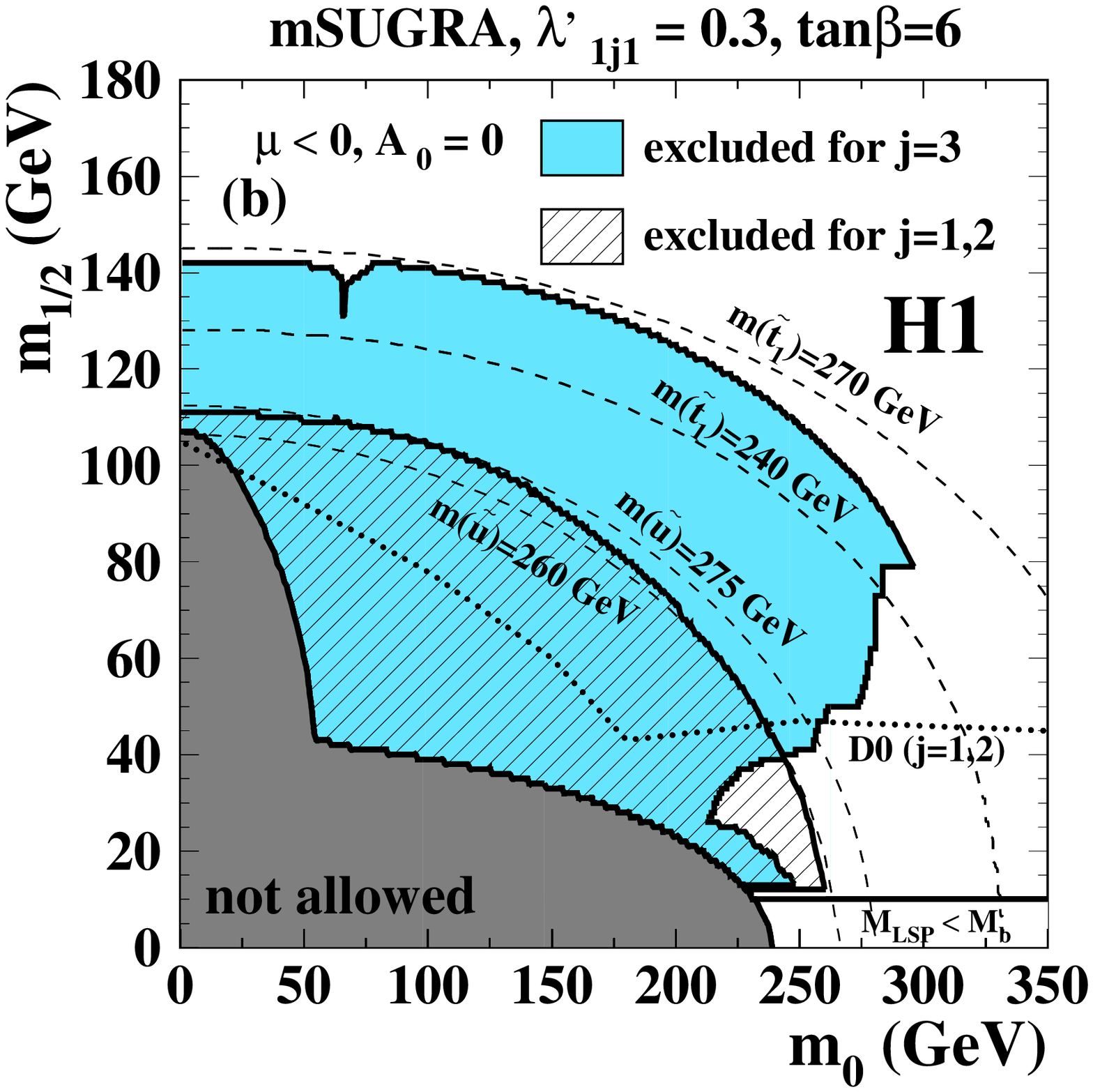,width=7.9cm}
  \end{center}
  \caption{
Excluded regions  ($95\,\%$ CL) in mSUGRA with $\lambda'_{1j1}=0.3$ for (a) 
$\tan\beta=2$ and (b) $\tan\beta=6$. The region marked ``not allowed'' corresponds 
to values of the 
parameters where no REWSB solution is possible or where the LSP is a sfermion.
The dashed lines indicate the curves of constant squark ($\tilde{u}_L$, $\tilde{t}_1$) mass.
The limits from LEP and the TeVatron are given by the dotted lines.
}
  \label{fig:sugraj}
\end{figure}

For $\mu < 0$, the exclusion limits at the $95\,\%$ CL obtained for a Yukawa coupling 
$\lambda'_{1j1} = 0.3$ ($j=1,2$) in the $(m_0, m_{1/2})$ plane are shown by the 
hatched histograms in figure~\ref{fig:sugraj} for the two example values (a) 
$\tan \beta = 2$ and (b) $\tan\beta=6$. The corresponding results for  
$\lambda'_{11k} = 0.3$ ($k=1,2$) are shown in figure~\ref{fig:sugrak}. The domains 
marked ``not allowed'' correspond to parameter values where no REWSB solution 
is possible or where the LSP is a sfermion.

The constraints on $(m_0, m_{1/2})$ are very similar for both values of $\tan\beta$ 
and both \Rp\ coupling types, $\lambda'_{1j1}$ and $\lambda'_{11k}$, since the mixing 
of the squark states is very small for $j=1,2$ and $k=1,2$. The excluded regions 
approximately follow curves of constant squark mass. For $\lambda'_{1j1} = 0.3$, the 
parameter space defined by {\mbox{$M_{\tilde{q}} < 275\,\GeV$}} is nearly fully excluded.
For $\lambda'_{11k} = 0.3$, the squark mass limit is slightly higher, 
{\mbox{$M_{\tilde{q}} < 285\,\GeV$}}, because of the higher squark production cross section 
in $e^-p$ collisions for equal couplings.

The results of the searches for \Rp\ SUSY by 
the D0 experiment~\cite{D0RES} at the TeVatron, 
which exploit di-electron events, are also 
shown in figures~\ref{fig:sugraj} and \ref{fig:sugrak}. 
For $\tan \beta = 2$, the H1 
limits are more stringent only for low values of $m_0$, whereas for $\tan \beta = 6$ 
the domain excluded by H1
extends considerably beyond the region ruled out by the D0 experiment.
For $\tan \beta =2$, the parameter space is more 
strongly constrained by the searches for 
$\chi$'s and sleptons at the L3 experiment~\cite{L3RPV} at LEP, as shown in 
figures~\ref{fig:sugraj} and \ref{fig:sugrak}. This is the only $\tan \beta$ value considered 
in \cite{L3RPV}. Results for higher values are expected to be similar. The LEP and TeVatron limits are independent of the Yukawa coupling.

\subsubsection{Results on stop and sbottom production}

A non-vanishing coupling $\lambda'_{131}$ would lead to the production of a stop squark. 
The weak stop eigenstates $\tilde{t}_L$ and $\tilde{t}_R$  mix through an angle 
$\theta_{\tilde{t}}$ to form the mass eigenstates 
$\tilde{t}_1=\cos\theta_{\tilde{t}}\, \tilde{t}_L+\sin\theta_{\tilde{t}}\, \tilde{t}_R$ 
and $\tilde{t}_2=-\sin\theta_{\tilde{t}}\,\tilde{t}_L+\cos\theta_{\tilde{t}}\,\tilde{t}_R$, 
whose production cross sections scale as $\lambda^{'2}_{131}  \cos^2\theta_{\tilde{t}}$ 
and $\lambda^{'2}_{131}  \sin^2\theta_{\tilde{t}}$, respectively.
Thus the lighter state $\tilde{t}_1$ does not necessarily have the largest production cross 
section. Similarly, for a non-vanishing $\lambda'_{113}$, sbottom production could be possible.
The weak sbottom states $\tilde{b}_L$ and $\tilde{b}_R$ mix to form the mass eigenstates 
$\tilde{b}_1=\cos\theta_{\tilde{b}}\,\tilde{b}_L+\sin\theta_{\tilde{b}}\,\tilde{b}_R$ 
and $\tilde{b}_2=-\sin\theta_{\tilde{b}}\,\tilde{b}_L+\cos\theta_{\tilde{b}}\,\tilde{b}_R$ 
and the production cross section for $\tilde{b}_1$ ($\tilde{b}_2$) scales as
$\lambda^{'2}_{113}  \sin^2\theta_{\tilde{b}}$ ($\lambda^{'2}_{113}  \cos^2\theta_{\tilde{b}})$. 
The treatment of stop production is described in the following.
Sbottom mixing is treated in the same way.

For the selection channels where the signal is integrated over the whole mass range,
the fraction of the visible signal  in a given selection channel $k$, is
$\sum_{i=1,2} (\varepsilon\beta)_{k,i}  \sigma_i / \sigma_{\rm tot}$,
where $(\varepsilon\beta)_{k,i}$ is the total visible branching ratio\footnote{The total 
visible branching ratio $(\varepsilon\beta)_k$ of a selection channel $k$ is given by 
$(\varepsilon\beta)_k=\sum_j \varepsilon_{k,j}\beta_j$, where $\beta_j$ is the branching 
ratio of the squark decay mode $j$ and  $\varepsilon_{k,j}$ is the corresponding efficiency 
in the selection channel $k$. The sum runs over all decay modes $j$ considered in the 
selection channel.} of the selection channel $k$ for  the state $\tilde{t}_i$, 
$\sigma_i$ is the production cross section of $\tilde{t}_i$ and 
$\sigma_{\rm tot} = \sigma_1 + \sigma_2$ is the total signal cross section.
For the channels in which the signal is integrated over a mass bin only the contribution 
of the state $\tilde{t}_i$ for which the sensitivity is maximal, {\it i.e.} for which 
$\sigma_i  (\sum_k (\beta\varepsilon)_{k,i} )$ is maximal, is taken into account in 
the above summation. The numbers of observed and expected events 
are then integrated in the mass bin corresponding to $\tilde{t}_i$ only.
\begin{figure}[t] 
  \begin{center}
    \epsfig{file=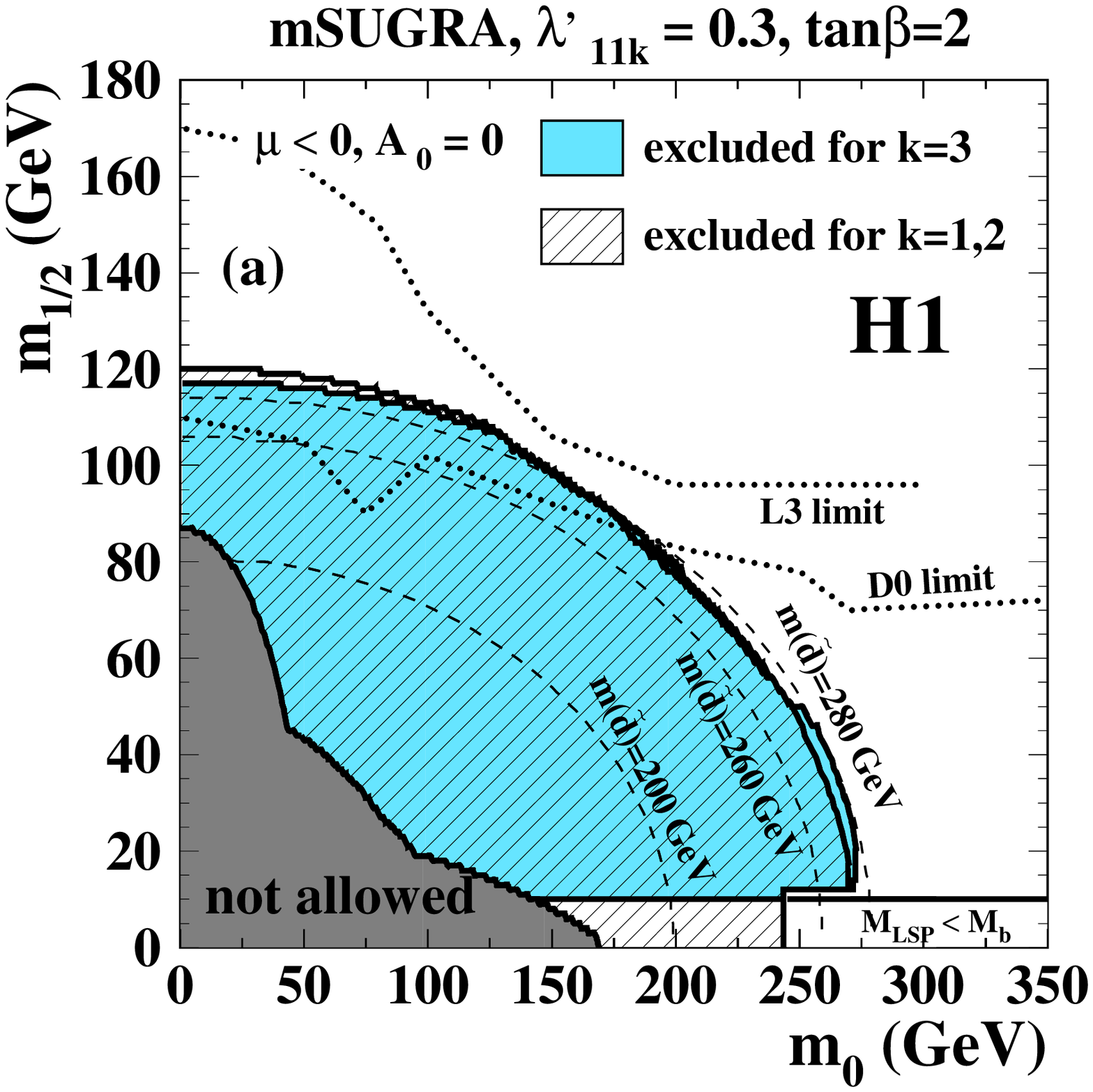,width=7.9cm}
    \epsfig{file=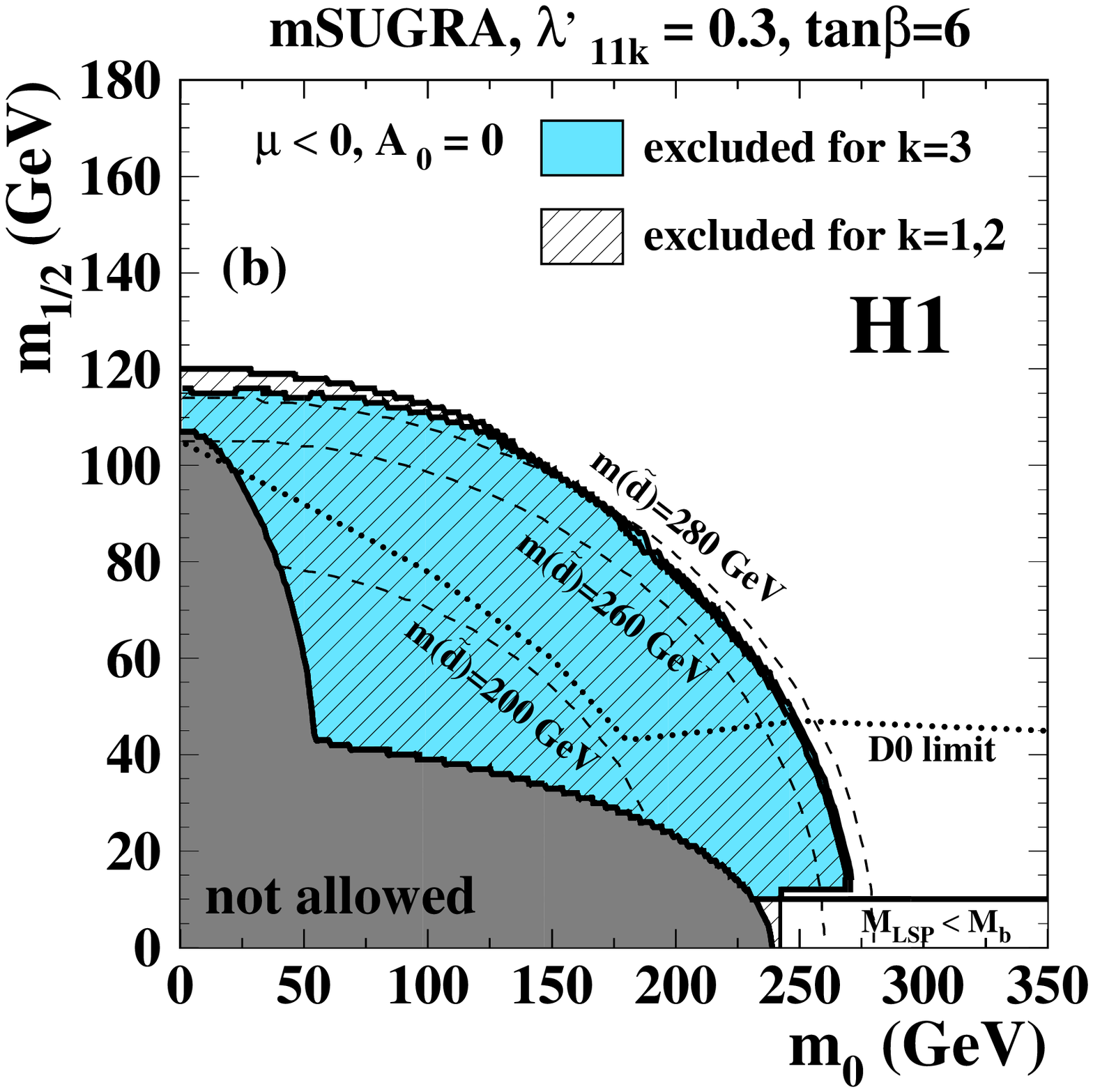,width=7.9cm}
  \end{center}
  \caption{
Excluded regions ($95\,\%$ CL) in mSUGRA with $\lambda'_{11k}=0.3$ for (a) $\tan\beta=2$ 
and (b) $\tan\beta=6$. The region marked ``not allowed'' corresponds to values 
of the parameters where no 
REWSB solution is possible or where the LSP is a sfermion. The dashed lines indicate the 
curves of constant squark mass. The limits from LEP and 
the TeVatron are given by the dotted lines.
}
  \label{fig:sugrak}
\end{figure} 

Using this procedure for both the stop and the sbottom case,  exclusion limits are derived 
for $A_0=0$ and $\mu < 0$ for $\tan \beta = 2$ and $\tan\beta=6$. The excluded regions in 
the $(m_0, m_{1/2})$ plane for $\lambda'_{131} = 0.3$ and $\lambda'_{113}=0.3$ are shown 
in figures~\ref{fig:sugraj} and \ref{fig:sugrak}, respectively. The domain below the line 
$m_{1/2} \simeq 10\,\GeV$ is not considered since it corresponds to cases where 
the only possible LSP decay, into $\nu b \bar{d}$, is kinematically forbidden.

In the case of stop production for $\tan \beta = 2$, shown in figure~\ref{fig:sugraj} 
(a), the excluded domain is slightly larger than that ruled out previously for 
$\lambda'_{1j1} = 0.3$ ($j=1,2$) due to the mixing in the stop sector which leads 
to $\tilde{t}_1$ masses smaller than the masses of the other squarks. Consequently, larger 
values of $m_{1/2}$ and $m_0$ can be probed. As shown in figure~\ref{fig:sugraj} (b) this 
remains the case for $\tan \beta = 6$ as long as $m_{1/2}$ is large enough to ensure 
that the mass of the lightest neutralino is above $30\,\GeV$. When the $\chi^0_1$ 
becomes too light, the detection efficiencies for the channels involving a $\chi_1^0$ 
(in particular the process $\chi^+_1 \rightarrow \chi^0_1$) are set to zero and
the sensitivity is only through the $eq$ channel or the decays
$\tilde{t} \rightarrow b \chi^+_1$ followed by a \Rp\ decay of the chargino. 
For even smaller $m_{1/2}$, if the $\chi_1^+$ mass is below $30\,\GeV$, only the $eq$ 
channel contributes. For $\tan \beta=2$ ($\tan\beta=6$), $\tilde{t}_1$ masses up to 
$265\,\GeV$ ($270\,\GeV$) can be excluded for $\lambda'_{131} = 0.3$.
These masses are smaller than the maximal sensitivity reached for the same coupling value 
 for $j=1,2$ because of the $\cos^2 \theta_{\tilde{t}}$ reduction of the $\tilde{t}_1$ 
cross section. For $\tan \beta=2$, the limits obtained from this analysis are comparable 
to the LEP sensitivity in $\chi$ and slepton searches at intermediate values of $m_0$.
In the same part of the parameter space, the H1 limits for higher values of $\tan\beta$ 
extend considerably beyond the LEP sensitivity which is expected to be similar to that 
for $\tan\beta=2$.

In the case of sbottom production (figure~\ref{fig:sugrak}) the limits are similar to 
those obtained for $k=1,2$, since the mixing in the sbottom sector is small at the 
$\tan\beta$ values considered. Thus, the mass difference between $\tilde{b}_1$ 
and $\tilde{b}_2$ is small. The sensitivity follows curves of equal $\tilde{b}_2$ 
masses because the production cross section for this state is much higher than 
for $\tilde{b}_1$ if the mixing angle is small. The parameter space leading to 
$\tilde{b}_2$ masses less than $280\,\GeV$ is ruled out.

\subsubsection{Dependence of the results on {\boldmath{$\tan\beta$}}}
\begin{figure}[t] 
  \begin{center}
    \epsfig{file=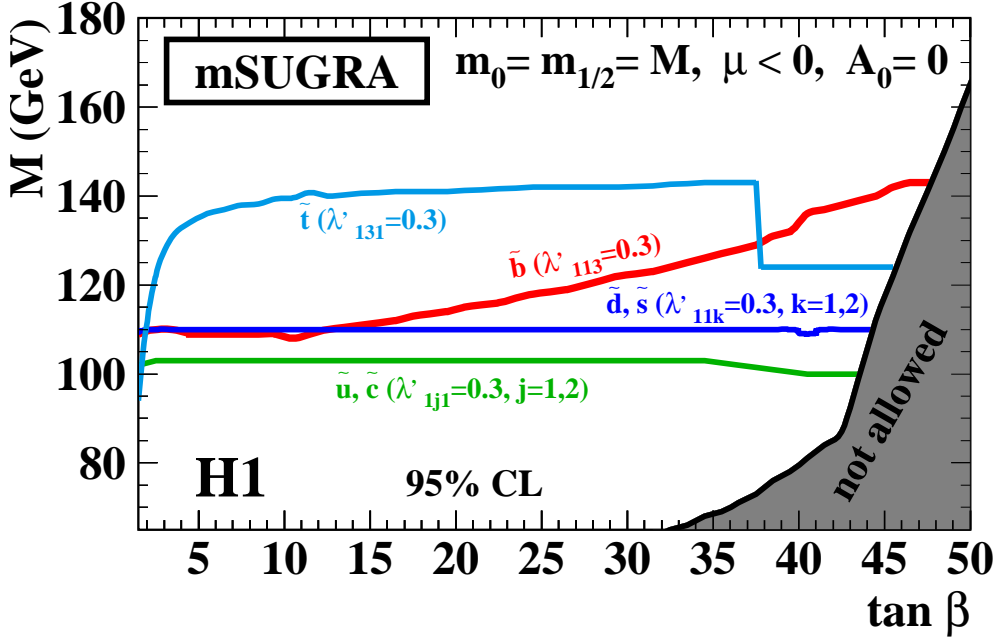,width=13.5cm}
  \end{center}
  \caption{Exclusion limits for $m_0=m_{1/2}=M$ in mSUGRA as a function of $\tan\beta$. 
The 95\,\% CL exclusion limits for $\lambda'_{1jk}=0.3$ are shown. The areas below the 
curves are excluded. The region marked ``not allowed'' corresponds to values 
of the parameters where no 
REWSB solution is possible or where the LSP is a sfermion.}

  \label{fig:msugratan}
\end{figure} 
In order to extend the parameter space to larger values of $\tan\beta$, 
a scan of this parameter is carried out. 
The number of free parameters is reduced by setting 
the masses $m_0$ and $m_{1/2}$ to a common value $M$. The 95\,\% CL limits 
on $M$ are shown in figure~\ref{fig:msugratan}  as a function of $\tan \beta$ 
for $\lambda'_{1jk}=0.3$. All squark flavours  are considered.
For the first two families the exclusion curves are rather flat since mixing 
effects are very small. Assuming equal $\Rp$ couplings, a larger part of the 
parameter space is excluded for $\tilde{d}$ and $\tilde{s}$ production than for 
$\tilde{u}$ and $\tilde{c}$ production because of the higher squark production 
cross section in $e^-p$ collisions. For squarks of the third family, mixing 
effects become important. For $\tan\beta\,\gsim\,10$ the increase of the mixing 
angle $\theta_{\tilde{b}}$ results in an improvement of the sbottom limit since 
it leads to a smaller $\tilde{b}_1$ mass, giving a higher $\tilde{b}_1$ production 
cross section. The mixing effects are largest in the stop sector, leading to more 
stringent limits on $M$. For very low values of $\tan\beta$, the 
$\cos^2\theta_{\tilde{t}}$  reduction of the $\tilde{t}_1$ production cross section 
is important. At values of $\tan\beta\,\,\gsim\,\,37$, the mixing of the two 
stau ($\tilde{\tau}$) states leads to decay chains involving light 
$\tilde{\tau}$s 
which result in final states including $\tau$ leptons. These channels are not 
searched for explicitly. Thus, in this region of the parameter space, the limit on 
stop production becomes less restrictive.  

\section{Conclusion}
A search for the \Rp\ production of squarks in $e^+ p$ and $e^- p$ 
collisions at HERA has been presented. No significant deviation from the SM is 
observed in any of the final state topologies resulting from direct or 
indirect 
$R_p$ violating squark decays. Mass dependent limits on the \Rp\ couplings 
$\lambda'_{1jk}$ are derived within a phenomenological version of the MSSM.
The existence of $\tilde{u}_L$-type and $\tilde{d}_R$-type squarks of all three 
generations with masses up to $275\,\GeV$ and $280\,\GeV$, respectively,
 is excluded at the $95 \%$ CL, for a Yukawa coupling equal 
to $\sqrt{4\pi\alpha_{\rm em}}$, in a large part of the MSSM parameter space.
These mass limits extend considerably beyond the reach of other collider experiments.
For lower squark masses, the results improve the indirect bounds set by low-energy 
experiments. Exclusion limits are also derived in the more restricted mSUGRA model, for 
which the limits obtained are competitive with and complementary to those derived at 
the LEP and TeVatron colliders.

\section*{Acknowledgements}
We are grateful to the HERA machine group whose outstanding efforts have made this 
experiment possible. We thank the engineers and technicians for their work in constructing 
and now maintaining the H1 detector, our funding agencies for financial support, the DESY 
technical staff for continual assistance and the DESY directorate for support and for the 
hospitality which they extend to the non-DESY members of the collaboration.
 
%
%

\clearpage

\end{document}